\documentclass[11pt,a4paper]{article}
\pdfoutput=1

\usepackage{amsmath,amsfonts,amsbsy,amssymb,enumerate,array}
\usepackage{graphicx,latexsym,verbatim,psfrag}
\usepackage[usenames]{color}
\usepackage[english]{babel}
\usepackage{fancybox}
\usepackage{slashed}
\usepackage{subfigure}
\usepackage{tcolorbox}%For the boxes in the conjectures

\usepackage{bm} %math bold symbols
\usepackage{booktabs} %some alignment stuff
\usepackage{chngpage} % allows for temporary adjustment of side margins

\usepackage{psfrag}

\usepackage{epsf}
\usepackage{amsthm}
\usepackage{xfrac}
\usepackage{tikz}
\usepackage[utf8]{inputenc}
\usepackage{cancel}

\usepackage{cite}   % in conjunction with below citation packages, combines citations into e.g. [4--7],
\usepackage{mciteplus}

\usepackage[colorlinks=true,      linkcolor=blue,      urlcolor=blue,
            filecolor=blue,      citecolor=blue,       pdfstartview=FitH,
						pdfpagemode=UseNone,      bookmarksopen=true]{hyperref}  % LINKS
\usepackage[all]{hypcap}     %should be loaded AFTER hyperref, which should otherwise be loaded last.

\usetikzlibrary{arrows,automata,positioning,calc,trees,decorations.pathmorphing,decorations.markings}

\newcommand\varpm{\mathbin{\vcenter{\hbox{%
  \oalign{\hfil$\scriptstyle+$\hfil\cr
          \noalign{\kern-.3ex}
          $\scriptscriptstyle({-})$\cr}%
}}}}

\newcommand\varmp{\mathbin{\vcenter{\hbox{%
   \oalign{\hfil$\scriptstyle-$\hfil\cr
           \noalign{\kern-.3ex}
          $\scriptscriptstyle({+})$\cr}%
}}}}

\def\be{\begin{equation}}
\def\ee{\end{equation}}
\def\bea{\begin{eqnarray}}
\def\eea{\end{eqnarray}}
\def\bal{\begin{align}}
\def\eal{\end{align}}

\def\mp{{M_{\text{P}}}}

\newcommand{\captionfonts}{\small}

\makeatletter  % Allow the use of @ in command names
\long\def\@makecaption#1#2{%
  \vskip\abovecaptionskip
  \sbox\@tempboxa{{\captionfonts #1: #2}}%
 \ifdim \wd\@tempboxa >\hsize
    {\captionfonts #1: #2\par}
  \else
    \hbox to\hsize{\hfil\box\@tempboxa\hfil}%
  \fi
  \vskip\belowcaptionskip}
\makeatother   % Cancel the effect of \makeatletter

\definecolor{cardinal}{rgb}{0.6,0,0}
\definecolor{darkgreen}{rgb}{0,0.4,0}
\definecolor{golden}{rgb}{0.92, 0.7, 0}
\definecolor{midnight}{rgb}{0, 0, 0.5}
\definecolor{darkblue}{rgb}{0, 0, 0.7}
\definecolor{boxblue}{rgb}{0.98, 0.98, 1.0}
\newcommand{\Red}{\color{red}}

\def\MG#1{{\bf \Red M:}}
\def\AH#1{{\bf \Red A:} }

\begin{document}

\numberwithin{equation}{section}

\begin{flushright}
\end{flushright}
\pagestyle{empty}
%\vspace{8mm}
\vspace{20mm}

\begin{adjustwidth}{-15mm}{-15mm} % to adjust the L and R margins
 \begin{center}
{\LARGE \textbf{{The Swampland Conjectures: \\
{\vskip 0.2cm}
A bridge from Quantum Gravity to Particle Physics}}}

\vspace{10mm} {\Large Mariana Gra\~na and Alvaro Herr\'aez}\\

\vspace{5mm}

Institut de Physique Th\'eorique, \\
Universit\'e Paris Saclay, CEA, CNRS, \\
Orme des Merisiers,  F-91191 Gif sur Yvette, France \\

\vspace{.5mm} {\small\upshape\ttfamily mariana.grana, alvaro.herraezescudero @ ipht.fr} \\

\vspace{15mm}
% \vspace{30mm}

%%%%%%%%%%%%%%%%%%%%%%%%
% Abstract
%%%%%%%%%%%%%%%%%%%%%%%%

\textbf{Abstract}
\end{center}
\end{adjustwidth}

\vspace{3mm}

\begin{adjustwidth}{6.5mm}{6.5mm} % to adjust the L and R margins
%\begin{center}
%\begin{quotation}
The swampland is the set of seemingly consistent low-energy effective field theories that cannot be consistently coupled to quantum gravity. In this review we cover some of the conjectural properties that effective theories should possess in order not to fall in the swampland, and we give an overview of their main applications to particle physics. The latter include predictions on neutrino masses, bounds on the cosmological constant, the electroweak and QCD scales, the photon mass, the Higgs potential and some insights about supersymmetry.

%\end{quotation}
%\end{center}
\end{adjustwidth}

%\thispagestyle{empty}
%\newpage

%\baselineskip=12.3pt
%\parskip=2pt

%\tableofcontents

%\baselineskip=15pt
%\parskip=3pt

\fontsize{11}{14}\selectfont

\newpage
\setcounter{page}{1}
	\pagestyle{plain}
\tableofcontents

\section{Introduction}

Recent years have seen the emergence of a new picture, or rather, a new paradigm, of quantum gravity. It has become clear that certain low energy theories that seem consistent from several points of view (such as e.g. anomaly cancellation) cannot be coupled to quantum gravity in a consistent way. 
The low-energy theories  that cannot be consistently coupled to gravity are said to belong to the \textit{swampland} \cite{swampland}.

On the other hand, the exploration of string theory has lead to a widely accepted picture of the space of string vacua. It is clear that string compactifications from ten to four dimensions may give rise to an enormous number of ground state solutions, commonly referred to as the \textit{landscape} of string theory. These vacua, which differ from each other in the shape and size of their six-dimensional internal spaces, actually give rise to a very rich set of \textit{universes} with completely different phenomenological properties. In particular, several of them  posses attractive  properties in that they come close to the Standard Model (SM) of particle physics or resemble interesting cosmological scenarios.

In the space of consistent low energy effective field theories (EFTs), the border separating the landscape from the swampland is delineated by a set of conjectures on the properties that these theories should have/avoid in order to allow a consistent completion into quantum gravity. Probably the first swampland conjecture is the statement that quantum gravity does not admit global symmetries \cite{Banks:2010zn}, which must be either gauged or broken at high energies. Other remarkable and more recent examples of such conjectures are the \textit{Weak Gravity Conjecture} \cite{ArkaniHamed:2006dz} and the (refined) \textit{de Sitter swampland conjecture} \cite{dS1,dS2,dS3} (see also \cite{Danielsson:2018ztv,Andriot:2018wzk}). Roughly speaking, the former states that outside of the swampland gravity must always be the weakest force, and the latter speculates that metastable de Sitter (dS) vacua belong to the swampland and therefore the nature of dark energy in our universe cannot be a cosmological constant.

All of these conjectures are supported by string theory arguments and examples, and many of them have also been widely studied in general holographic setups (see e.g. \cite{Harlow:2018jwu, Harlow:2018tng,Montero:2018fns}). Furthermore, most of them originally arise and can also be heuristically motivated by thinking about the physics of black holes, in particular about black hole decays and black hole remnants.\footnote{Even though along this work we always have string theory in mind when referring to quantum gravity, these conjectures are believed to hold in any possible consistent theory of quantum gravity with weakly-coupled Einstein gravity as its low-energy limit, mainly because this kind of black hole arguments can equally be applied to those setups.} For example, the conjecture about the absence of global symmetries in quantum gravity is based on the fact that otherwise an infinite number of black hole remnants with masses of the order of the Planck mass could form and render the theory out of control, as explained in more detail in section \ref{ss:globalsymmetries}. Unfortunately, not all of them can be stated in such simple terms, and their evidence mainly comes from exploring string theory constructions and pinpointing general patterns. Moreover, the number, connections among, and consequences of the swampland conjectures has increased significantly over the last five years, leading to a change of paradigm: the theories that can be consistently coupled to quantum gravity are highly constrained.  This is not such a preposterous idea: EFTs always break down at some cutoff scale, and in order to go beyond this scale one should include new degrees of freedom. The new degrees of freedom should be such that they do not lead to inconsistencies in a quantum theory of gravity. If the UV complete theory is to describe quantum gravity, black holes should inevitably be part of the spectrum, thus bringing in all their intricate nature.   

The goal of this review is to give a summary of the main implications of the Swampland Program for particle physics, explaining the logic behind. 
There are also a lot of implications for cosmology coming from the Swampland Program, but we will not cover them here. It is of course difficult to draw a sharp boundary between cosmology and  particle physics, so to be precise we will not discuss problems such as the quantum consistency of dS vacua or early universe cosmology and inflation (we refer the interested reader to the reviews \cite{Brennan:2017rbf,Palti:2019pca,vanBeest:2021lhn} and references therein).

Let us remark  that most of the implications are ``postdictions" instead of predictions, in the sense that they allow to understand some of the quantities that have been measured instead of predicting them. Still, it is remarkable that quantum gravity considerations has strong consequences  for particle physics. 
%Still, this is clearly a huge step forward, as highly reducing the parameter space of solutions while  still obtaining that the remaining part is compatible with observations is quite a non-trivial task, especially coming from quantum gravity considerations. In fact, it is good to keep in mind that this is actually the way in which progress has been typically made in physics and science in the past, namely measuring observables and then explaining them (this is the case for example of the spin of the electron, which was first measured experimentally and found to be half-integer, and then explained within the framework of QFT). 
Moreover, we are only at the first stages of constraining phenomenology from quantum gravity, but the expectation is that as we understand better the swampland conjectures, they will become more predictive and provide concrete guidance for phenomenology.

Some of the main open problems in particle physics these days are neutrino masses, the nature of the Higgs boson and the structure of the Higgs potential, the electro-weak hierarchy problem, the strong CP problem, the existence of supersymmetry and its associated breaking scale and, of course, the big elephant in the room of theoretical physics: the cosmological constant. As we will see, the Swampland Program can provide interesting insights into these problems, as well as unexpected relations between them. In particular, it is remarkable how things that seem not to be related from a low energy EFT point of view can be mysteriously connected in quantum gravity, as for example the value of neutrino masses and that of the cosmological constant. In this sense, there is a key aspect that seems to need a reformulation once quantum gravity enters the game, namely that of naturalness:
%. To be a bit more precise,  the most important lesson of the Swampland Program is that not everything that is naively allowed from the point of view of low energy EFTs is actually allowed in quantum gravity. With this in mind, 
properties that may seem \emph{unnatural} from the point of view of the parameter space of EFTs might be perfectly \emph{natural} from the point of view of quantum gravity. Despite the fact that this might sound surprising, this idea of UV/IR mixing has been explored in the literature (see e.g. \cite{Cheung:2014vva,Lust:2017wrl,Craig:2018yvw,Craig:2019fdy}). In fact, we will present evidence that some of the apparently strange hierarchies that appear in our universe, and that would be very unnatural from an EFT point of view, can be related among themselves through quantum gravity arguments, hence alleviating some of these naturalness issues. 

As an example, take the mass of the Higgs boson. As a scalar, at the EFT level radiative corrections \emph{naturally} render its mass of the order of the UV cutoff. However, as we will see, quantum gravity arguments generically predict the cutoff to be lower than what one would naively expect (e.g. smaller than $M_P$). Moreover, we will also present some swampland arguments that relate the Higgs vev to other (a priori unrelated) parameters of the theory, which are intrinsically gravitational, such as the cosmological constant scale. These relations between different quantities that are completely independent and disconnected at the low energy EFT level, but which arise once quantum gravity arguments are considered, are precisely what should be taken into account for a \emph{reformulation of naturalness}. In some sense, this is the main lesson from the Swampland Program, namely that the naive EFT parameter space will drastically  be reduced when Quantum Gravity is included in the game. Thus, some region of that naive parameter space that may seem unnatural from the EFT point of view, may be the only one left in the presence of gravity, and therefore be completely natural.

The strcuture of this review is as follows. The swampland conjectures that have an implication in particle physics are reviewed in section \ref{sec:conjectures}, while its consequences are explained in section \ref{sec:implicationsPP}. Some final discussion is presented in \ref{sec:discussion}.

\section{The Swampland Conjectures}
\label{sec:conjectures}

In this section we present the set of swampland conjectures that are relevant for the phenomenological applications discussed in section \ref{sec:implicationsPP}. This is by no means a comprehensive review of all the conjectures, but the selected ones are presented in a logical way, such that the main ideas and evidence for them, as well as some of their underlying relations, can be followed consistently. We refer the interested reader to the reviews \cite{Brennan:2017rbf,Palti:2019pca,vanBeest:2021lhn} for more exhaustive and technical presentations. 

\subsection{Absence of Global symmetries and Cobordisms}
\label{ss:globalsymmetries}

The absence global symmetries in quantum gravity is generally considered as the first swampland conjecture. Unlike other conjectures, it is hard to give credit to a unique paper or author proposing it in the first place, but the standard reference is \cite{Banks:2010zn}. The conjecture can be stated as follows.
\vspace{0.2cm}

\begin{tcolorbox}[colback=boxblue]
\textbf{No-Global Symmetries Conjecture}: There cannot be exact global symmetries in a theory of quantum gravity coupled to a finite number of degrees of freedom.
\end{tcolorbox}

First of all, let us recall that a global symmetry is, roughly speaking,  a transformation that commutes with the Hamiltonian of a theory and transforms (some subset of the) physical states into different physical states (i.e. it acts non-trivially on the Hilbert space).\footnote{For a useful and extended discussion about the precise definition of global symmetries, as well as a proof of their absence in the context of holography, see \cite{Harlow:2018jwu,Harlow:2018tng}} A first motivation for the absence of global symmetries in string theory is the fact that global symmetries on the worldsheet are actually gauged in from the point of view of target space \cite{Banks:1988yz}. However, at this point a natural question may arise: What about the global symmetries of the Standard Model? In particular, there is no experimental signature of the breaking of the global B-L symmetry of the SM so far, so are we saying that the SM is in the swampland? The answer is no, and the reason clarifies why it is difficult to extract phenomenologically interesting constraints from the No-global symmetries conjecture. The key point is that only \emph{exact} global symmetries are forbidden in quantum gravity. Therefore it is perfectly fine to have a low energy theory with an apparent global symmetry, as long as it is either gauged or broken at high-energies. The obstruction to obtain meaningful phenomenological constraints from this conjecture is precisely the fact that we cannot say anything about the scale at which this gauging or breaking must take place. However, it is still extremely interesting to consider this conjecture for several reasons. The first is the fact that it can still give us very useful information to understand the fundamental principles behind quantum gravity, which is a perfectly valid motivation per se. Furthermore, some other conjectures that we will introduce latter, such as the Weak Gravity Conjecture or the Swampland Distance Conjecture, can be morally seen as refinements of the idea of forbidding exact global symmetries in quantum gravity. 

Probably the simplest instance in which one can see how something goes wrong with an exact global symmetry in quantum gravity is to consider the case of a global $SU(2)$ \cite{OoguriCERN}. States in that theory will then be classified in irreducible representations of the group, which we label $j$ and whose dimension is given by $2j+1$.  Consider now a Schwarzschild black hole with mass $M$ and  horizon radius is $r_H=2 M/M_P^2$. By combining an appropriate number of particles charged under the non-trivial representations of $SU(2)$, we could then construct such a black hole with arbitrarily large $j$ (and possibly a very large mass). Nothing seems wrong up to this point, but let us consider the effect of Hawking evaporation on the black hole. In particular, since the symmetry is global, it means that there is no preferred charge carried by the particles that are emitted by Hawking radiation and hence the black hole will loose its mass but not its charge. This means that as the black hole loses its mass we would reach a contradiction for the entropy of such an object, since the Bekenstein Hawking entropy will not be able to accommodate a big enough dimension for the Hilbert space, which is required to account for the global symmetry. More concretely, we would reach the point for which
\begin{equation}
\mathrm{Dim}(\mathcal{H}_{\mathrm{BH}})\sim e^{S_{BH}} < (2j+1), 
\end{equation}
where $S_{BH}=A/4G_N$ is precisely the Bekenstein-Hawking entropy, which decreases with the mass of the black hole via its dependence in the area. This argument is particularly transparent, and serves the purpose of illustrating the kind of black hole arguments behind some of the swampland conjectures. 

The previous argument can also be adapted to abelian groups (or abelian subgroups of non-abelian groups). In a theory with a $U(1)$ global symmetry, for every black hole with a global charge $Q$ there will eventually be a remnant black hole at the final stage of Hawking evaporation, of mass $M \sim M_P$. This would then lead to an infinite number of remnants in a finite mass range, labeled by the different values of $q$. Such an infinite number of remnants within a finite mass range has been argued to drive the theory out of control by forcing the renormalized Planck mass to diverge \cite{Susskind:1995da}).

The No-global symmetries Conjecture can be extended to the case of discrete global symmetries and also to generalized global symmetries, where charged objects are not point-like but extended \cite{Gaiotto:2014kfa,Heidenreich:2020pkc,Heidenreich:2021tna}. It is of particular interest to consider the compactification of a $d$-dimensional theory on a $k$-dimensional compact manifold. Each such $k$-dimensional compactification space would give rise to a different $(d-k)$-dimensional  vacuum. One can then consider the existence of domain walls that connect the different vacua of such a $(d-k)$-dimensional theory. If there were disconnected (families of) vacua, namely, vacua that are not connected by any (finite energy) domain wall in the spectrum, one could construct a $(d-k-1)$-form global symmetry that would, therefore, violate the No-global Symmetries Conjecture. This lead to the proposal of the following conjecture 
\vspace{0.2cm}
\begin{tcolorbox}[colback=boxblue]
\textbf{Cobordism Conjecture} \cite{McNamara:2019rup}: The cobordism class of any $k$-dimensional compact space on which a $d$-dimensional theory of quantum gravity can be compactified must be trivial, i.e.
\begin{equation}
\Omega_k^{\mathrm{QG}}=0
\end{equation}
\end{tcolorbox}

\begin{figure}[t]
	\begin{center}
		\subfigure[]{	
			\includegraphics[scale=0.9]{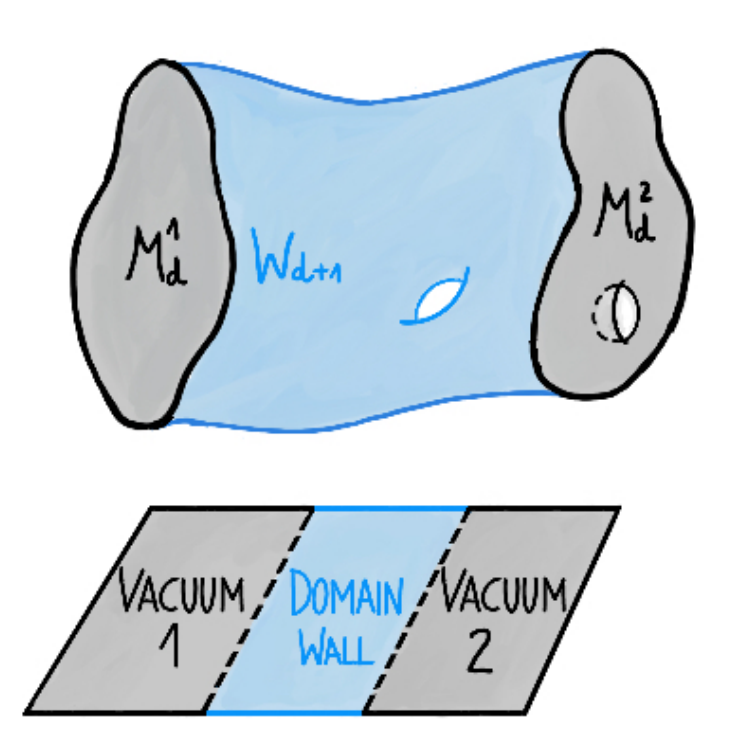} 
		    \label{fig:cobordantmanifolds}
		}
		\qquad \quad 
		\subfigure[]{
			\includegraphics[scale=0.9]{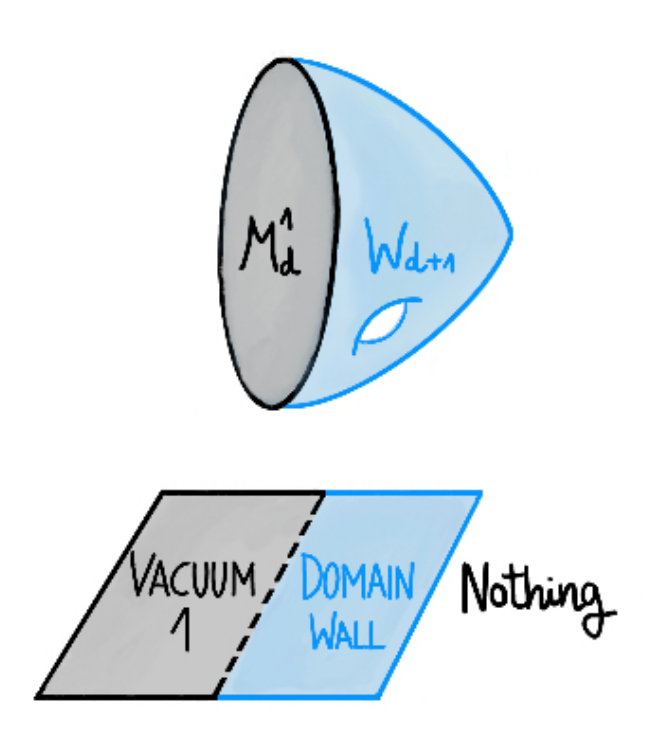}
			\label{fig:trivialcobordism}
		}
		\caption{\footnotesize \textbf{(a)} Two cobordant $d$-dimensional manifolds, which at the level of a lower dimensional EFT give rise to a domain wall separating a theory compactified on both of them. \textbf{(a)} A $d$-dimensional manifold in the trivial cobordism class. Compactifying a theory in such a manifold thus supports an end of the world membrane separating it from \emph{nothing}.}
	\end{center}
\end{figure}

Two $d$-dimensional manifolds, $\mathcal{M}_d^1$ and $\mathcal{M}_d^2$, are said to be cobordant if their union is the boundary of another $(d+1)$-dimensional manifold, $W_{d+1}$. From the point of view of a lower dimensional vacuum obtained from compactification of a theory on $\mathcal{M}_d^1$, the fact that it is cobordant to $\mathcal{M}_d^2$ can be interpreted as a domain wall separating the original vacuum and the one obtained by compactifying the same theory in $\mathcal{M}_d^2$, as shown in Fig. \ref{fig:cobordantmanifolds}. Furthermore, a manifold is in the trivial cobordism class if it is a boundary itself and compactifying a theory on such a manifold there exists a domain wall that acts as an end of the world membrane, separating the vacuum from \emph{nothing}, as displayed in Fig. \ref{fig:trivialcobordism}. 

Let us briefly outline the argument presented in \cite{McNamara:2019rup} to construct a global charge for a $(d-k-1)$-defect  in the presence of a non-trivial cobordism group by considering gravitational solitons (see \cite{Witten:1985xe}). Consider on the one hand the theory on $d$-dimensional Minkowski space $\mathcal{M}_d$, and on the other hand a consistent compactification on a $k$-dimensional space denoted by $X_k$. One can always construct a gravitational instanton in the $d$-dimensional theory by removing a small $k$-dimensional ball both from $M_d$ and from $X_k$ and then gluing them together. From the $d$-dimensional theory point of view, this can be seen as a $(d-k-1)$-dimensional defect. Moreover, if the cobordism class of $X_k$ were non-trivial, it would be impossible to deform the space in the presence of such gravitational instanton to the original $M_d$ space upon time evolution. Therefore, there would exist an invariant of $X_k$ which would would not change under time evolution. This is precisely the definition of  a global charge. We could then think of throwing such $(d-k-1)$-dimensional defects into $(d-k-1)$-dimensional black brane solutions, which would then carry such a global charge, and would yield the same problems as we already explained. It is therefore necessary to have $\Omega_k^{\mathrm{QG}}=0$ in order to avoid the existence of such global charges in quantum gravity. As a last remark, let us mention that the global charge we are considering here is of topological nature,\footnote{A simple example of a topological charge is the skyrmion number of solitonic solutions in theories with a global $SU(2)$ symmetry (see e.g. \cite{Tong:GaugeTheories}).} as opposed to the global charges coming from continuous global symmetries, but the same black hole/brane arguments introduced above, as well as other evidences in favor of the absence of global symmetries in quantum gravity, apply equally well to both types of charges.

%The absence of global symmetries gives rise to the conjecture that the cobordism class of quantum gravity theory should be trivial \cite{McNamara:2019rup}. 
In general, this conjecture implies that in a $k$-dimensional compactification space, there must exist configurations describing a $(k+1)$-dimensional geometry whose boundary is the $k$-dimensional space. Any $k$-dimensional space is then viewed as an end of the world defect. For example, the Horava-Witten boundary is the cobordism defect of 11d M-theory, while the O8-planes are the defect of type IIA theory. According to this conjecture, the other 10d superstring theories should also admit cobordism branes, but these are unknown, and are expected to be fairly exotic (and non-supersymmetric).

\subsection{The Weak Gravity Conjecture and beyond}

The Weak Gravity Conjecture (WGC) was originally  proposed in \cite{ArkaniHamed:2006dz}, and it has received an enormous amount of attention in the recent years (see for instance \cite{Brown:2015iha,Rudelius:2015xta,Heidenreich:2015nta, Heidenreich:2015wga,Ibanez:2015fcv,Marsh:2015xka,Hebecker:2015zss, Heidenreich:2016aqi, Montero:2016tif,Saraswat:2016eaz,McAllister:2016vzi,Brandenberger:2016vhg,Palti:2017elp,Montero:2017mdq,Ibanez:2017vfl, Aldazabal:2018nsj,Lee:2018urn,Lee:2018spm, Andriolo:2018lvp,Cheung:2018cwt,Heidenreich:2019zkl,Lee:2019tst, Gonzalo:2019gjp,Benakli:2020pkm,Benakli:2020vng, Buratti:2020kda, Gonzalo:2020kke} for an incomplete list of  works related to the Weak Gravity Conjecture and its applications to particle physics and cosmology) resulting in several generalizations and refinements, as we will comment shortly. In its original formulation, it includes two claims, the so-called electric and magnetic versions, which state the following:\footnote{We focus here in the 4d version. Note that $\mathcal{O}(1)$ factors and powers of the Planck mass would change in different dimensions.}
\vspace{0.2cm}
\begin{tcolorbox}[colback=boxblue]
Given a gravitational theory with a $U(1)$ gauge symmetry with gauge coupling $e$, 
\begin{itemize}
\item[]{\textbf{Electric Weak Gravity Conjecture} \cite{ArkaniHamed:2006dz}: The spectrum of the theory must include at least a particle with mass $m$ and charge $q$ that satisfies the inequality 
\begin{equation}
\label{eq:EWGC}
m^2 \leq 2 e^2 q^2 M_{P}^2\, .
\end{equation}}
\item[]{\textbf{Magnetic Weak Gravity Conjecture} \cite{ArkaniHamed:2006dz}: There exists an upper bound for the UV cutoff of the EFT, given by
\begin{equation}
\label{eq:MagWGC}
\Lambda_{\mathrm{UV}} \lesssim e M_P\, .
\end{equation}
}
\end{itemize}
\end{tcolorbox}

\noindent Let us explain the extent of these conjectures. For concreteness, let us consider Einstein-Maxwell theory, with action
\begin{equation}
\label{eq:EinsteinMaxwell}
S_{E-M}=\int d^{4} x \sqrt{-g}\left(\frac{M_{P}^{2}}{2} R-\frac{1}{4 e^{2}} F_{\mu \nu} F^{\mu \nu}+\ldots\right)
\end{equation}
where $g$ is the determinant of the metric, $R$ is the corresponding Ricci scalar, the field strength is given by $F_{\mu \nu}=\partial_{[\mu} A_{\nu]}$ and the ellipses indicate any possible extra coupling to matter. In order to build the kinetic term for a field $\phi$ with quantized charge $q \in \mathbb{Z}$ one uses the covariant derivative
\begin{equation}
D_\mu \phi= \left(\partial_\mu + i q A_\mu \right) \phi \, .
\end{equation}
The global part of this gauge symmetry acts on the field as $\phi \rightarrow e^{2\pi i q \alpha} \phi$, with $\alpha$ a constant parameter. In particular, if we take $e\rightarrow 0$ we recover an exact global symmetry, since the kinetic term for the gauge bosons diverges and they decouple, so that only the global part of the symmetry remains. From this argument  it is clear that the Weak Gravity Conjecture is deeply related to the aforementioned absence of global symmetries in quantum gravity, since its magnetic version states that the cutoff of the theory would go to zero in the limit in which the global symmetry is restored. 

The heuristic black hole arguments supporting the Weak Gravity Conjecture go along the same lines as the ones previously introduced to motivate the absence of global symmetries in quantum gravity. The relevant black holes to consider in this case are of Reissner-Nordstr\"om  type, which arise as black hole solutions to the action \eqref{eq:EinsteinMaxwell}. They are characterized by their mass, $M$, and their (quantized) charge, $Q$. Their so-called extremality bound is given by
\begin{equation}
M^2 \geq 2 e^2 Q M_{P}^2.
\end{equation}
If this inequality is satisfied, the black hole has two horizons and it is called subextremal. If it is saturated, the two horizons coincide and the black hole is dubbed extremal. Finally, the third possibility, namely that the inequality is not fulfilled, yields a superextremal black hole. This final case has no horizon and would thus describe a naked singularity, which is known to be problematic according to cosmic censorship \cite{Penrose:1969pc}. 
One can build different subextremal black holes with arbitrarily large charge by making their masses larges enough. Now if one considers only Hawking radiation in the context of Einstein-Maxwell theory, it turns out that the black holes evaporate loosing both their mass and charge until they approach the extremal limit, so that if we wait long enough there will be extremal black holes for every value of $Q$ with mass $M  \, \sim \, e \, Q M_{P}$, where we are assuming $Q$ is big enough for the semiclassical calculation to remain valid. If all these extremal black holes were stable, in the weak coupling limit (i.e. $e \rightarrow 0$) there would be an arbitrarily large number of extremal black hole states for a finite mass range, yielding similar problems to those encountered in the presence of infinitely many stable remnants. An economical solution to make these extremal states unstable and avoid this problem is to require the existence of a (super)extremal particle, which would allow these extremal black holes to decay. 
%The requirement that the particle be superextremal is a purely kinematic statement, since it is a necessary condition for the black hole to be able to emit such a particle and become subextremal, and it was one of the original motivations for the WGC \cite{WGC}

As one can imagine, there are some subtleties and potential loopholes in these arguments and a rigorous proof  is still an open question. 
%For instance, they rely on the validity of the semiclassical computation of black hole evaporation until the extremal situation is reached. 
However, they serve to illustrate important points and can be applied quite generally. In particular, the requirement that extremal black holes should be able to decay can be generalized to black brane solutions, as done in \cite{Heidenreich:2015nta}, where a generalization of the Weak Gravity Conjecture for $p$-branes in $d$-dimensions was proposed by requiring the existence of branes fulfilling
\begin{equation}
      T^{2} \, \leq \, \gamma e^2 \,  q^{2} M_P^{d-2}\, .
    \label{eq:WGCqforms}
\end{equation}
Here, $T$ is the tension of the brane, $q$ its quantized charge and $e$ is the corresponding $(p+1)$-form gauge coupling. $\gamma$ is an $\mathcal{O} (1)$ factor that fixes the concrete extremality bound for the corresponding black brane solutions, and it can depend on $d$, $p$ and the details of the particular model under study. For large codimension objects, namely strings (codimension 2) or membranes (codimension 1) in 4d, this extension is somewhat tricky due to the strong backreaction effects, but similar bounds apply as discussed in detail in \cite{Lanza:2020qmt,Lanza:2021qsu}. 

Finally, let us mention a complementary way to approach the Weak Gravity Conjecture, also suggested in the original work \cite{ArkaniHamed:2006dz}, which emphasizes the importance of preventing the formation of fully stable gravitational bound states. If more and more bound states could be formed in the theory one would encounter problems which are similar to those caused by remnants, rendering it out of control. In this spirit, the basic form of the electric Weak Gravity Conjecture can be obtained by requiring that there exists a charged particle on which gravity acts as the weakest force. Taking the action \eqref{eq:EinsteinMaxwell}, the force between two objects of charge $q$ and mass $m$ separated by a distance $r$ includes an attractive contribution from the gravitational interaction and a repulsive one from the electromagnetic field. They are given by 
\begin{equation}
F_{\text {grav}}=\frac{m^{2}}{8 \pi M_{P}^{2} r^{2}}, \qquad F_{\mathrm{EM}}=\frac{(e q)^{2}}{4 \pi r^{2}},
\end{equation}
and the electric version of the Weak Gravity Conjecture can be stated as the requirement that the electromagnetic repulsion be stronger than the gravitational attraction, avoiding the formation of stable bound states. As in the previous case, these arguments can be generalized to extended objects to recover the condition \eqref{eq:WGCqforms} for (at least) a $p$-brane in the spectrum of the theory.

\subsubsection*{The Non-susy AdS Conjecture}
\label{ss:non-susyAdS}

\begin{figure}[t]
	\begin{center}
		\subfigure[]{	
			\includegraphics[scale=0.62]{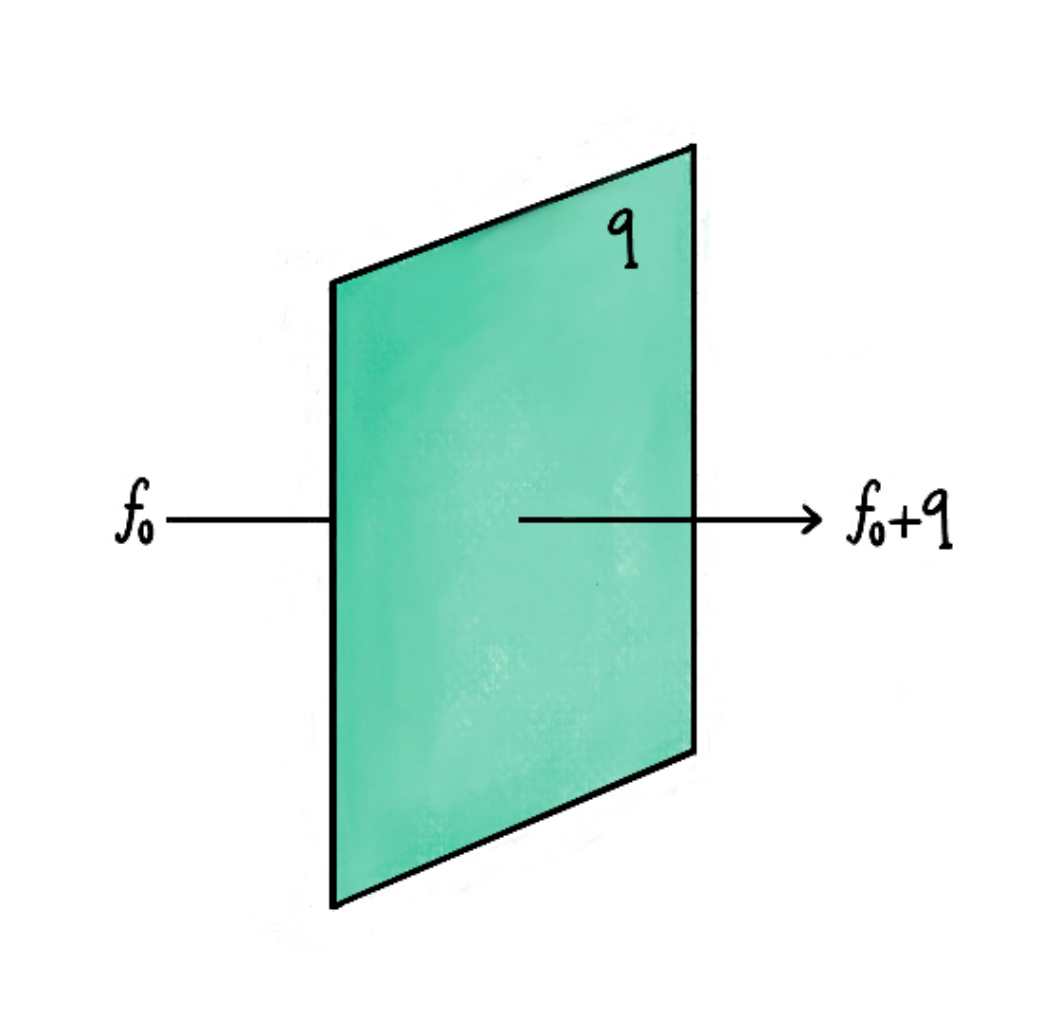} 
		    \label{fig:membranecrossing}
		}
		\qquad \quad 
		\subfigure[]{
			\includegraphics[scale=0.62]{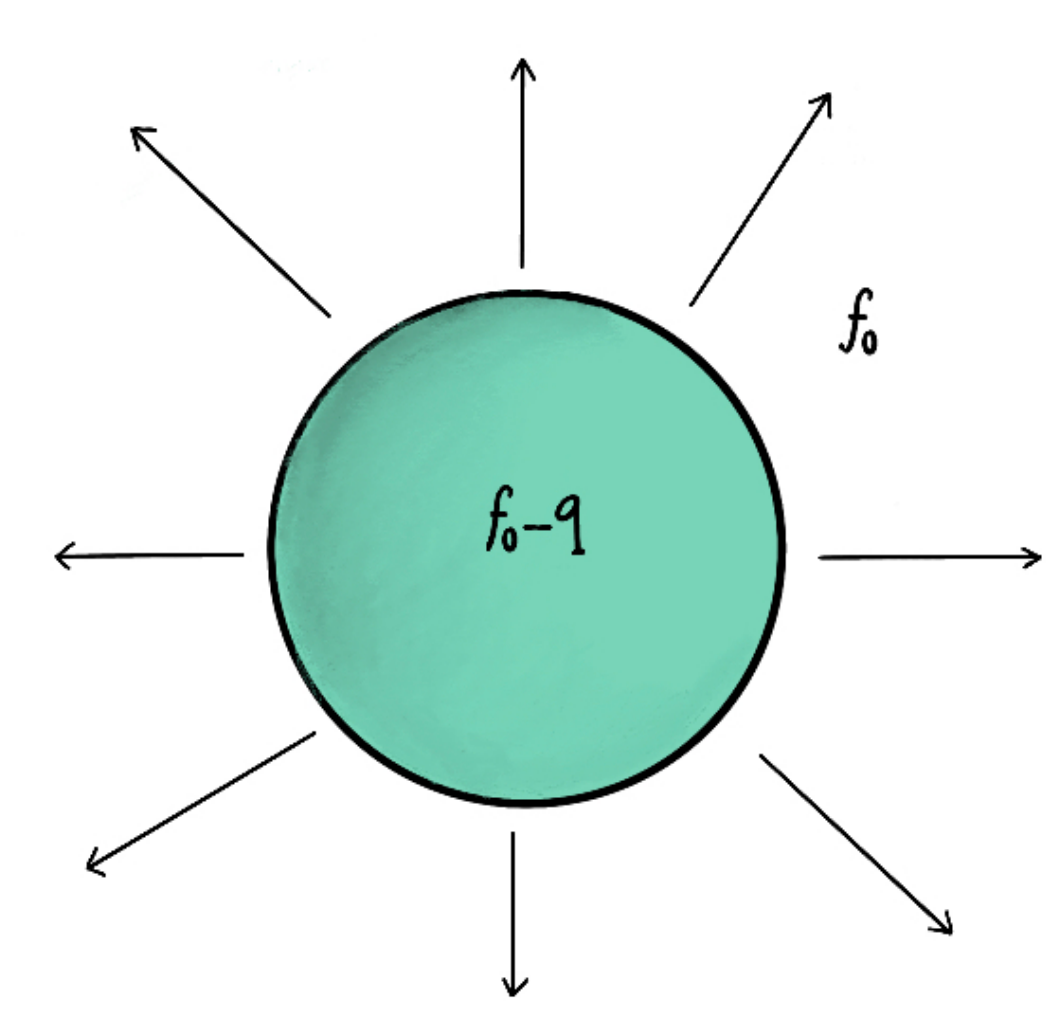}
			\label{fig:bubbletransition}
		}
		\caption{\footnotesize \textbf{(a)} Shift in a flux $f_0$ as a consequence of crossing a membrane with charge $q$ under the dual 3-form. \textbf{(b)} Vacuum decay by nucleation of a bubble that expands and is surrounded by a membrane with charge $q$, such that the flux outside the bubble, $f_0$, is reduced in the internal region of the bubble.}
	\end{center}
\end{figure}

In this section we present a swampland Conjecture which has particularly interesting consequences for low energy physics, as we will explain in section \ref{sec:implicationsPP}. To do so, let us first focus on the typical stringy setup to obtain lower dimensional vacua, which includes the presence of internal fluxes for the different $p$-form field strengths in the spectrum. That is, the solutions admit (and typically require) non vanishing values for $\int_{\Sigma_p} F_p$, where $\Sigma_p$ is a non-trivial $p$-cycle in the internal manifold. From the point of view of the 4d effective theory, these non-vanishing internal fluxes can be seen to be dual to 4-form field strengths, whose corresponding 3-form gauge fields are the objects that  naturally couple to  (codimension 1) membranes.  In fact, the charge of such membranes, $q_{\mathrm{mem}}$, with respect a 3-form corresponds to the jump in the flux dual to the 4-form, as displayed in Fig. \ref{fig:membranecrossing}. Vacuum decay can then proceed via the nucleation of a bubble formed by such a membrane, which would separate spacetime into two regions (two vacua) with different values for the fluxes. If the electric repulsion of the walls of such a bubble overcomes their gravitational attraction, the bubble will then expand and the original vacuum would decay to the vacuum inside the bubble. Otherwise the nucleated bubble collapses and the original vacuum would be stable under such a decay.

Having briefly introduced flux vacua, the other crucial ingredient in order to understand the motivation behind the \emph{non-susy Anti-de Sitter conjecture} is  the refinement of the Weak Gravity Conjecture originally proposed in \cite{Ooguri:2016pdq}. It states that the Weak Gravity Conjecture inequality displayed in $\eqref{eq:EWGC}$, or more generally $\eqref{eq:WGCqforms}$, can only be saturated by BPS states in supersymmetric theories. In particular, when applied to (codimension 1) membranes in non-supersymmetric vacua, this implies the existence of  a superextremal one satisfying strictly
\begin{equation}
      T_{\mathrm{mem}}^{2} \, < \, \gamma e^2 \,  q_{\mathrm{mem}}^{2} M_P^{d-2}\, .
    \label{eq:WGCmemstrict}
\end{equation}
Furthermore, in Anti-de Sitter (AdS) vacua supported only by fluxes, it was shown in \cite{Maldacena:1998uz} that the presence of a strictly superextremal membrane always allows for the construction of a bubble separating a region with one less unit of flux from the original flux vacuum. Since this bubble grows, it mediates the transition from the initial vacuum to another one with one less unit of flux, as shown in Fig. \ref{fig:bubbletransition}. Therefore this refinement of the Weak Gravity Conjecture immediately implies that all non-susy AdS vacua supported only by fluxes must be unstable. This lead the authors of \cite{Ooguri:2018wrx} to conjecture this to be a general feature of quantum gravity:
\vspace{0.2cm}
\begin{tcolorbox}[colback=boxblue]
\textbf{Non-susy Anti-de Sitter Conjecture} \cite{Ooguri:2018wrx}: Any non-supersymmetric AdS vacuum in quantum gravity must be unstable.
\end{tcolorbox}

\begin{figure}[t]
	\begin{center}
		\subfigure[]{	
			\includegraphics[scale=0.45]{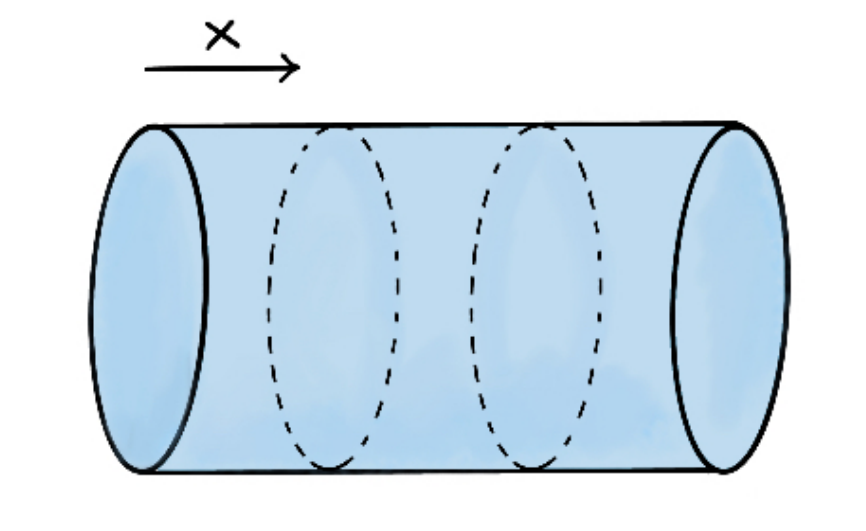} 
		    \label{fig:S1dimension}
		}
		\qquad \quad 
		\subfigure[]{
			\includegraphics[scale=0.45]{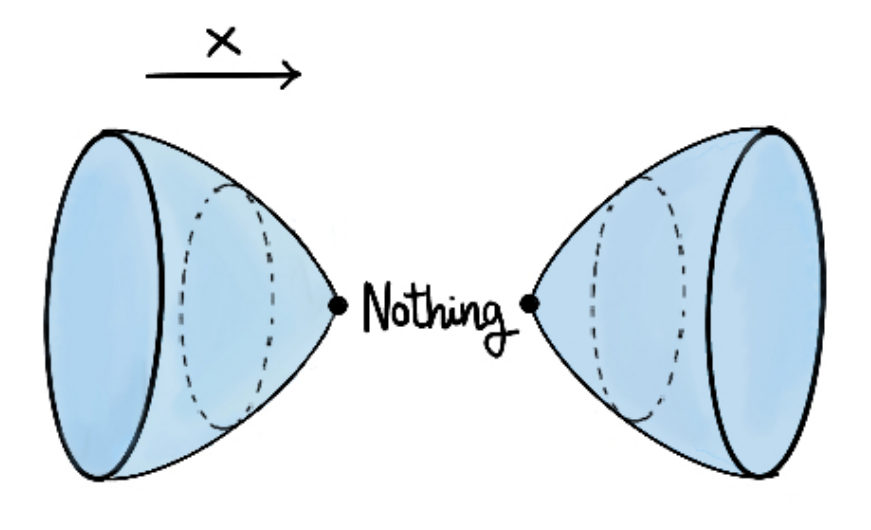}
			\label{fig:boncircle}
		}
		\subfigure[]{
			\includegraphics[scale=0.7]{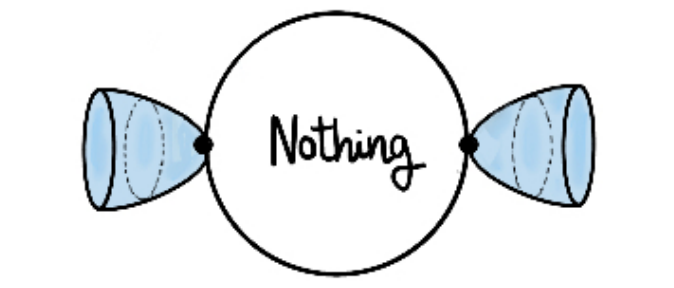}
			\label{fig:bon3d}
		}
		\caption{\footnotesize \textbf{(a)} Illustration of an extra circular dimension, the $x$-direction represents the non-compact space. \textbf{(b)} If  the size of the extra dimension (i.e. the circle) is reduced along the non-compact direction until it eventually reaches zero size, a region of space is removed (the one that corresponds to the values of $x$ that are between the two black dots). \textbf{(c)} A bubble of nothing in more dimensions is obtained when the size of the extra dimension becomes zero at the boundary of such a region. Here we have represented this schematically only at two points in the boundary of the bubble, but in reality there would be one cigar-like blue figure attached to each point on the boundary of the bubble.}
		\label{fig:BONS1}
	\end{center}
\end{figure}

Let us finally introduce an extra motivation for this conjecture relating it to the Cobordism Conjecture. Probably the first question that comes to mind after the previous discussion is what kind of universal channel may be responsible for the decay of more general AdS vacua (e.g. with scalar fields sourced by the fluxes, without fluxes...). In \cite{GarciaEtxebarria:2020xsr} (see also \cite{Dibitetto:2020csn}) it was argued that possible candidate for this might be a \emph{bubble of nothing}. Bubbles of nothing, originally introduced in \cite{Witten:1981gj}, are non-perturbative instabilities that arise in theories with compact extra dimensions in which spacetime  can decay into nothing. This might seem bizarre, but it is easier to visualize in the original example treated in \cite{Witten:1981gj}, namely a circle compactification. This is displayed in figs. \ref{fig:S1dimension}-\ref{fig:boncircle}, where it can be seen that the non-compact direction ends if the compact circle collapses to zero size, so that some part of space that was originally there disappears, but the resulting geometry is perfectly smooth and complete. With more than one non-compact dimension, some region of space can collapse to nothing if the size of the circle vanishes at the boundary of such a region, as shown schematically in Fig. \ref{fig:bon3d}, where the circle represents the boundary of that region (the bubble of nothing) and at every point of it the size of the extra dimension is zero. In the presence of extra ingredients, such as fermions, there may be topological obstructions for the circle to be shrunk to zero size. This is indeed the case in the presence of periodic boundary conditions for fermions on the circle, and therefore the bubble of nothing is only allowed with anti-periodic boundary conditions. 

For more general $d$-dimensional compact manifolds, the question of whether they can collapse to zero size depends precisely on  whether they belong to the trivial cobordism class,  as this would mean that the internal manifold is itself the boundary of a $(d+1)$-dimensional manifold. The Cobordism Conjecture thus ensures that bubbles of nothing are always topologically allowed in quantum gravity. It is important to remark though that the Cobordism Conjecture does not directly imply the Non-susy AdS Conjecture, as the absence of a topological obstruction does not mean that it is always dynamically favorable for a bubble of nothing to expand and mediate the decay of the vacuum. It could happen that the dynamics of such a bubble force it to shrink and disappear instead of expand, or even to form infinite flat domain walls. Still, even though the dynamics of the bubble are hard to solve in general, there are indications that the conditions for it to expand and mediate vacuum decay are closely related to the breaking of supersymmetry \cite{GarciaEtxebarria:2020xsr}. This, together with the aforementioned argument using the refined version of the Weak Gravity Conjecture conjecture for membranes, constitute some solid evidence supporting the Non-susy AdS Conjecture.

\subsubsection*{The Festina Lente bound}

In this and the previous section we have introduced  the black hole arguments that originally lead to some of the best established swampland conjectures, namely the absence of global symmetries and the Weak Gravity Conjecture. It seems reasonable to say that the take away message is that (sub)extremal black holes must be able to decay while remaining (sub)extremal, as otherwise the theory  gives rise to several problematic scenarios such as the troubles with remnants or naked singularities commented before. In particular, avoiding the appearance of such naked singularities (i.e. Weak Cosmic Censorship) plays a crucial role in the Minkowski case presented above and it is also related to the Weak Gravity Conjecture in AdS \cite{Crisford:2017zpi,Crisford:2017gsb,Horowitz:2019eum}. Furthermore, taking it as a guiding principle and applying it to dS space, as originally done in \cite{Montero:2019ekk} (and further extended in \cite{Montero:2021otb}), turns out to give rise to a new bound for the spectrum of theories that do not belong to the swampland. In the following, we briefly present the arguments that lead to this bound and leave the discussion about its main phenomenological implications to section \ref{ss:FLpheno}.

Consider the 4d Einstein-Maxwell theory given by the action \eqref{eq:EinsteinMaxwell}  in a dS background with cosmological constant $\Lambda_{\mathrm{dS}}=3 M_P^2/ \ell^2_{\mathrm{dS}}$ (with $\ell_{\mathrm{dS}}$  the dS length)\footnote{Note that in our conventions the cosmological constant has units of mass to the power $d$, as opposed to the usual conventions taken in General Relativity where it is set to have units of mass squared by extracting explicitly a factor of $M_P^{d-2}$.}
\begin{equation}
\label{eq:EinsteinMaxwelldS}
S=\int d^{4} x \sqrt{-g}\left(\frac{M_{P}^{2}}{2}  R-\Lambda_{\mathrm{dS}}  -\frac{1}{4 e^{2}} F_{\mu \nu} F^{\mu \nu}+\ldots\right).
\end{equation}

\begin{figure}[t]
	\begin{center}
		\subfigure[]{	
			\includegraphics[scale=0.26]{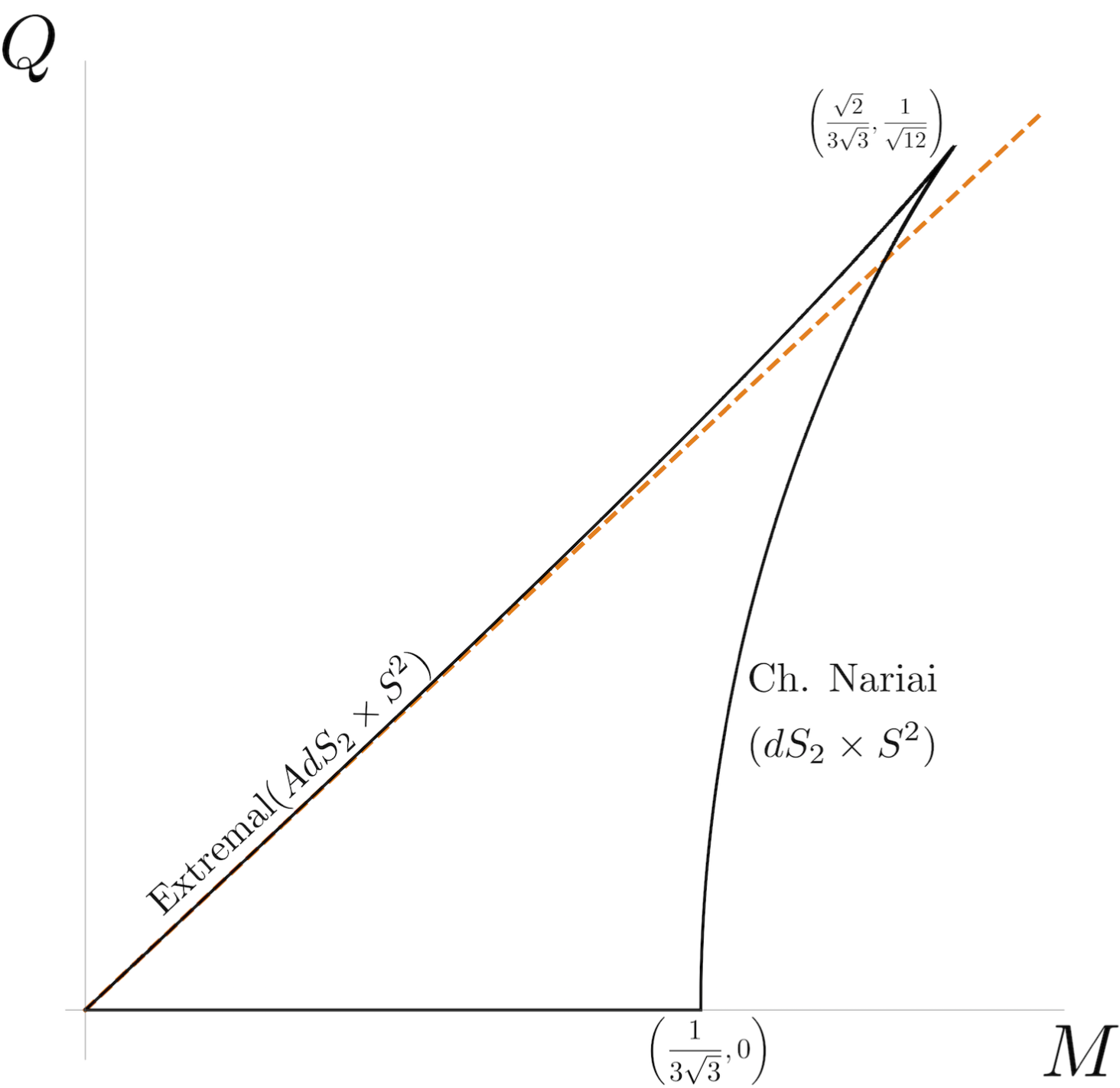} 
		    \label{fig:sharkfin}
		}
		\qquad \quad 
		\subfigure[]{
			\includegraphics[scale=0.3]{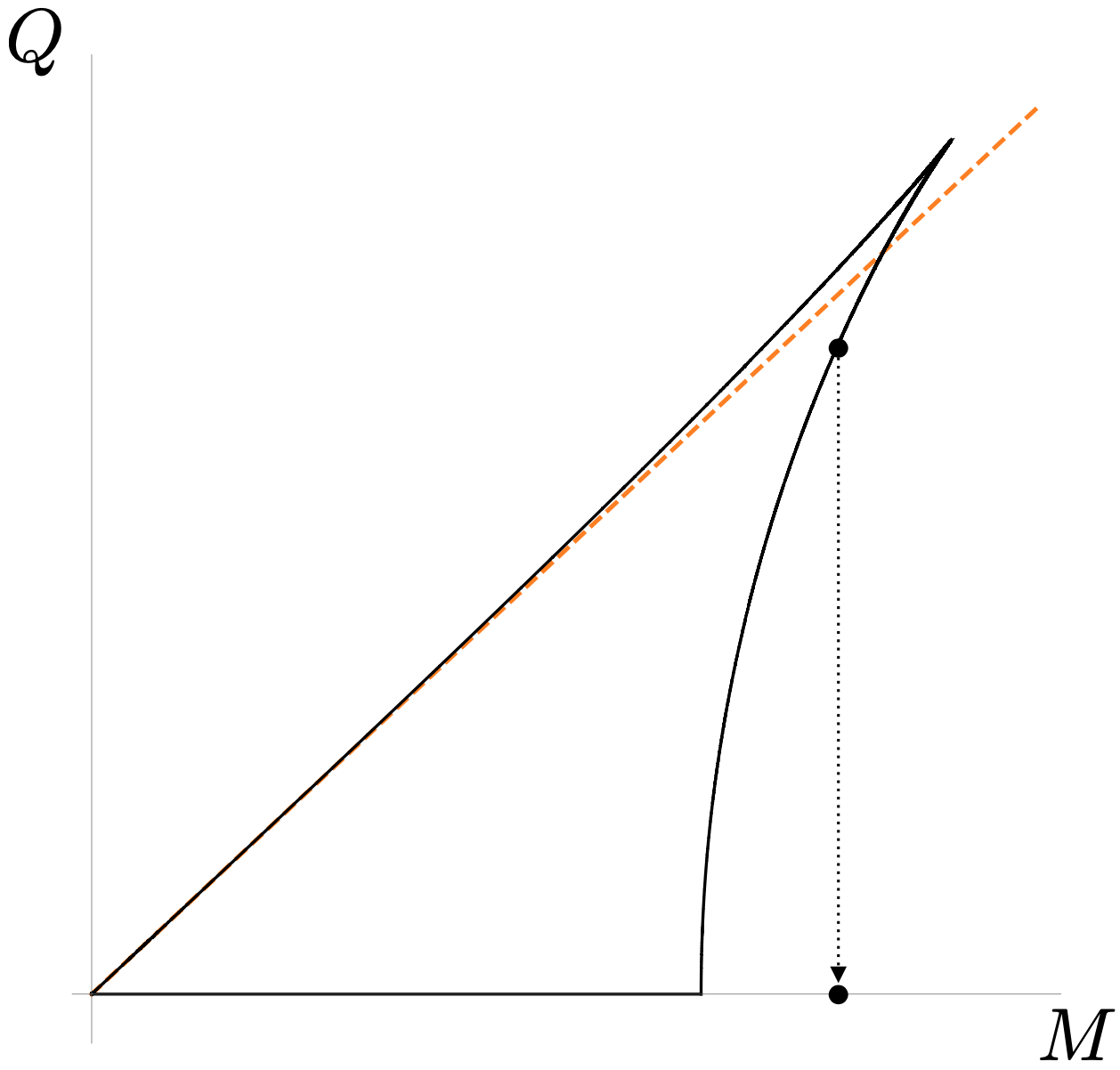}
			\label{fig:adiabaticdecay}
		}
		\caption{\footnotesize  Figs. taken from \cite{Montero:2019ekk}. \textbf{(a)} Phase space of  Reissner-Nordstr\"om-de Sitter black holes. \textbf{(b)}Decay of Nariai black holes by emission of  particles with $m^2\ll qE$, so that they discharge almost instantaneously without effectively loosing mass.}
	\end{center}
\end{figure}

We are interested in the charged black hole solutions to this action, the so-called Reissner-Nordstr\"om-de Sitter black holes. These solutions generically have three horizons, the two that are the analogous to the Reissner-Nordstr\"om solution in flat space, plus the cosmological horizon of dS. Their phase space is shown in Fig. \ref{fig:sharkfin}. The usual extremality bound, given by the coincidence between the first two horizons, corresponds to the upper line and solutions above it are superextremal. Additionally, there is an upper bound for the mass of a black hole for a given charge, which arises due to the fact that the black-hole horizon must remain within its own dS cosmological horizon. The limiting case, where both horizons coincide, corresponds to the so-called charged Nariai black holes, depicted as the curved line in the right of the figure. These solutions are a key difference in dS, and study of their decay in detail leads to the announced bound. Finally, the point where both the extremality line and the charged Nariai line meet is the so-called ultracold black hole, and it gives the highest values for the charge and the mass. The inside region thus contains all the possible subextremal charged black holes in dS.

The decay of black holes within the allowed region is expected to eventually end up in empty dS space according to the thermal picture of dS in the static patch. Roughly speaking, it states that one can interpret the physics in the static patch as a finite-dimensional thermodynamical system with temperature, so that any excitation of the dS vacuum corresponds to pushing the system out of equilibrium. The thermodynamical system eventually goes back to equilibrium (i.e. empty dS). 

Studying the decay of such black-holes by means of Hawking and Schwinger radiation has two interesting limits. In the so-called quasi-static limit, in which the produced particles have $m^2 \gg\ q E$ ($E$ is the electric field that supports the charged black hole solution), points within the allowed region slowly evolve towards the origin, finally ending in empty dS space, as expected.\footnote{As a side comment, this can be seen as one of the reasons why it is harder to argue for the usual Weak Gravity Conjecture in dS space. In dS, the mere existence of a charged particle is enough to allow for the black hole decay, since such particle can be pair produced at sufficiently long distance from the black hole for the dS expansion to overcome the gravitational attraction, independently of its charge (note still that this process can be extremely slow). By contrast, in Minkowski this attraction after pair production can only be compensated if the electric repulsion is large enough, giving rise to the usual Weak Gravity Conjecture constraint.} On the other hand, in the so-called adiabatic regime, where the emitted particles have $m^2 \ll \ q E$, the electric field is almost instantaneously screened by pair production and any charged Nariai black hole looses its charge without loosing mass, as displayed in Fig. \ref{fig:adiabaticdecay}. This is problematic for several reasons. First, it gives rise to a big crunch, instead of empty dS space. This is clearly in contradiction with the thermal picture of dS introduced before, unless some exotic unknown process made the big crunch solution transition back to empty dS, which sounds very unlikely. Second, the parameter space of black holes outside the allowed region is connected in a smooth way, so that one can easily deform the solution after decay, without crossing any of the black lines, to one above the extremality region (i.e. to a superextremal black-hole) which would again give rise to a naked singularity. Demanding that these inconsistencies do not show up is equivalent to demanding that the black-hole cannot decay in the adiabatic regime, and this is achieved if every particle in the spectrum fulfils $m^2 \gg \ q E$. In particular, this must be the case for all the allowed values of the electric field, which is maximum at the ultracold point and equal to $E=\sqrt{6}e M_P  H $, with $H$ the Hubble constant. Therefore one arrives to
\vspace{0.2cm}
\begin{tcolorbox}[colback=boxblue]
\textbf{Festina Lente Bound} \cite{Montero:2019ekk}: For a gravitational EFT on a dS background, every charged particle in the spectrum must satisfy
\begin{equation}
\label{eq:FLbound}
m^{4} \gtrsim 6\left(e q M_{P} H\right)^{2}=2(e q)^{2} V.
\end{equation}
\end{tcolorbox}

First of all, note that in the last step we have rewritten the bound in terms of a potential $V$. This would reduce to the corresponding cosmological constant in an purely dS background, but already in \cite{Montero:2019ekk} it was argued that all the arguments above would also apply to quintessence-like scenarios  (where instead of a cosmological constant there is a runaway potential) as long as they are sufficiently flat. More recently, in \cite{Montero:2021otb}, this has been studied in detail and the precise conditions that the potential must fulfil have been given.

Moreover, note that this is fundamentally different to  the usual Weak Gravity Conjecture in several ways. First, it gives a lower bound for the masses of the particles, instead of an upper one (it is a strong gravity condition in fact!). Second, the bound is significantly stronger in the sense that it must be fulfilled by every single charged particle in the theory, not only by a subset of them as required in the usual Weak Gravity Conjecture. This stronger nature of the constraint can be understood from the fact that in the arguments for the Weak Gravity Conjecture, it is enough to ensure that the black holes can decay in some way, whereas here one needs to avoid every possible decay in the adiabatic limit, not only some of them. Yet one more intriguing point is that, whereas in the flat limit, i.e. $H \rightarrow 0$, the bound becomes trivial (as expected since Nariai black holes are a purely dS phenomenon), the same does not happen when one naively takes the decoupling limit of gravity $M_P \rightarrow \infty$, with $e$ and $H$ fixed. In general it is expected that swampland constraints become trivial in the field theory limit, so this could be hinting towards an obstruction to taking the naive decoupling limit in dS. We will not elaborate further on this here, but refer the interested reader to \cite{Montero:2019ekk, Montero:2021otb} for a more detailed discussion.

Finally, let us mention that this bound stands on less rigorous grounds than other swampland conjectures where a lot of string theory evidence is known, mainly because no explicit dS construction fully under control has been found (yet?) in string theory. Still, the heuristic black hole arguments displayed here, together with the more complete and rigorous discussions in \cite{Montero:2019ekk, Montero:2021otb} are solid evidence for the conjecture. Furthermore an extra motivation has recently been put forward in \cite{Montero:2021otb}, where it was argued that the Festina Lente bound can be understood as a spacelike version of cosmic-censorship, forbidding crunches that are not hidden behind black-hole horizons. All this makes definitely interesting to consider its potential phenomenological consequences.

\subsection{Towers of states and the Distance Conjectures}

\label{sec:towers}

The Swampland Distance Conjecture (SDC) \cite{Ooguri:2006in} is one of the most studied and well-established swampland Conjetures (see \cite{Baume:2016, Klaewer:2016kiy,Blumenhagen:2017cxt, Blumenhagen:2018nts,Grimm:2018ohb,Grimm:2018cpv, Corvilain:2018lgw, Lee:2019xtm,Lee:2019wij,Blumenhagen:2019qcg,Marchesano:2019ifh,Font:2019cxq,Grimm:2019wtx,Baume:2019sry, Gendler:2020dfp,Baume:2020dqd,Perlmutter:2020buo, Bastian:2020egp,Klaewer:2020lfg,Calderon-Infante:2020dhm} for an incomplete list of recent works), and it introduces an omnipresent feature in EFT of quantum gravity, namely the appearance of infinite towers of states that become light an imply a breakdown of the EFT. In fact, the appearance of such light towers of states has been used as a rationale to extend and generalize the Swampland Distance Conjecture in several ways, as we will introduce below. In its original form, this conjecture can be stated as follows.
\vspace{0.2cm}
\begin{tcolorbox}[colback=boxblue]
\textbf{Swampland Distance Conjecture} (SDC) \cite{Ooguri:2006in}: Consider a gravitational effective theory with a moduli space (i.e. a space parameterized by the massless scalar fields in the theory) and whose metric is given by the kinetic terms of the scalar fields. Starting from a point $P$ in moduli space and moving towards a point $Q$ an infinite geodesic distance away (i.e.  $d(P,Q)\rightarrow \infty$), one encounters an infinite tower of states which become exponentially massless with the geodesic distance, i.e. 
\begin{equation}
\dfrac{M(Q)}{\mp}\ \sim \ \dfrac{ M(P)}{\mp} \ \, e^{-\alpha \,  d(P,Q)},
\end{equation}
with $\alpha$ an order one constant in Planck units. 
\end{tcolorbox}

It has also been proposed, that the exponential behaviour cannot be delayed by more than order one distances (in Planck units) and that the Swampland Distance Conjecture should also be satisfied in the presence of a scalar potential (as long as it does not obstruct the infinite distance points) \cite{Baume:2016, Klaewer:2016kiy}.

To give some intuition about the Swampland Distance Conjecture, let us consider the canonical example, namely a theory compactified on a circle of size $R$. It is well known that the Kaluza-Klein (KK) modes in such a circle compactification have a mass that scales with the internal radius as
\begin{equation}
\label{eq:KKmasesS1}
m_{n}^2=\dfrac{n^2}{R^2}, \qquad n \in \mathbb{Z}.
\end{equation}
After dimensional reduction of the gravitational piece of the action and the corresponding  field redefinition to go to the Einstein-frame, the kinetic term for the radion field $R$ takes the form $\mathcal{L}_{\mathrm{kin}}\sim \dfrac{\partial_M R \, \partial^MR}{R^{2}}$.
The distance between two points, $R_i$ and $R_f$  in field space is therefore measured by the field space metric, given by $1/R^2$, and it yields
\begin{equation}
d(R_i, R_f)=\int_{R_i}^{R_f} \sqrt{\dfrac{1}{R^2}}\, dR= \left|\log (R_f/R_i) \right|.
\end{equation}
Therefore, starting at any finite radius, there are two points that lie at infinite proper distance, namely $R\rightarrow \infty$ and $R\rightarrow 0$. Approaching the former it is clear that rewriting the KK tower masses given by eq. \eqref{eq:KKmasesS1} one obtains precisely the behaviour predicted by the Swampland Distance Conjecture, namely
\begin{equation}
m_{n}\sim e^{-d(R_i, R_f\rightarrow \infty)}.
\end{equation} 
On the other hand, approaching the infinite distance point $R\rightarrow 0$, one could be tempted to say that the Swampland Distance Conjecture is violated. However, this is not the case if we consider string theory, which includes an infinite tower of winding states with masses given by
\begin{equation}
m_{w}^2=\dfrac{w^2R^2}{(\alpha^\prime)^2}.
\end{equation}
These  become exponentially light in terms of the field-space distance $d$ as $R\rightarrow 0$ limit. Hence, string theory provides a natural candidate for the tower in the two possible limits. In fact, this also suggests a fundamental connection between the Swampland Distance Conjecture and the existence of extended objects in quantum gravity, as the winding states only appear when strings are considered. 

The Swampland Distance Conjecture can be understood as a restriction on the range of validity of any EFT coupled to gravity, in the sense that an EFT defined at a point in moduli space cannot be extended to a point which is at an arbitrarily large distance from the initial one. If one tried to do so, an infinite number of light degrees of freedom would become like and break the aforementioned EFT description. As in all swampland conjectures, this is to be compared to the situation in which gravity is not present, in which no obstruction to the extension of an EFT to an arbitrary point in moduli space appears. A neat microscopic interpretation for the Swampland Distance Conjecture is not fully clear at the moment, but it is strongly inspired by dualities in string theory. In the KK example above this picture is indeed realized by T-duality. Along these lines, it has been conjectured that every infinite distance limit actually corresponds to either a decompactification limit or a string becoming tensionless, hence having a dual theory with a different object as the fundamental string in that limit \cite{Lee:2019xtm, Lee:2019wij}. 

Finally, let us also mention that the aforementioned towers, whose energy scale is related to the breaking of the EFT, fit very naturally with the picture presented by the Weak Gravity Conjecture, particularly with its magnetic version. This is the case because weak-coupling points are generally at infinite distance in moduli space. The lower and lower cutoff scale predicted by the magnetic Weak Gravity Conjecture as we approach those limits may then be associated with the presence of a tower of states, which actually motivated the proposal of the so-called tower versions of the Weak Gravity Conjecture. These are the \emph{(Sub)lattice Weak Gravity Conjecture} \cite{Heidenreich:2015nta,Heidenreich:2016aqi}, which requires the existence of a superextremal particle (i.e. a particle fulfilling eq. \eqref{eq:EWGC} or its higher-dimensional generalization \eqref{eq:WGCqforms} ) at every point in a (sub)lattice of the lattice of charges, and the \emph{Tower Weak Gravity Conjecture} \cite{Andriolo:2018lvp}, which predicts \eqref{eq:EWGC} the existence of an infinite number of superextremal particles, not necessarily populating a sublattice.  In fact, It is well known that in many examples in string theory, the states in the tower that satisfy the Swampland Distance Conjecture are also the states that satisfy tower versions of the Weak Gravity Conjecture. The possibility of this being a result of the restoration of a global symmetry at every infinite distance point was suggested in \cite{Gendler:2020dfp}. This is indeed the case in the circle compactification presented above, where the tower of KK states are charged under the $U(1)$ graviphoton and saturate the Weak Gravity Conjecture bound. Moreover, the winding modes are charged under the 1-form coming from the reduction of the $B$-field along the circle, and they also saturate the Weak Gravity Conjecture inequality.

%The Anti-dS Distance Conjecture can be seen as a particular case of a generalized version of the Swampland Distance Conjecture, namely the Generalized Distance Conjecture \cite{lpv}. 

The usual formulation of the Swampland Distance Conjecture deals with the distance on the space of scalar fields (i.e. the moduli space). Nonetheless, this notion of distance can be generalized to a notion of distance between more general field configurations, applicable to any (tensor) field  with a generalized metric given again by its kinetic terms \cite{Lust:2019zwm}. The claim of the \emph{Generalized Distance Conjecture} \cite{Lust:2019zwm} is then that an infinite tower of states becomes light exponentially with this generalized distance as it diverges.\footnote{Of course, this distance reduces to the usual distance in moduli space when only massless scalars are varied, and the Generalized Distance Conjecture reduces to the usual Swampland Distance Conjecture}  In particular, when it applied to families of vacua with different values for the cosmological constant, it gives rise to the following conjecture.
\vspace{0.2cm}
\begin{tcolorbox}[colback=boxblue]
\textbf{Anti-de Sitter Distance Conjecture} (ADC) \cite{Lust:2019zwm}: In a $d$-dimensional theory of quantum gravity with cosmological constant $\Lambda_d$, there exist a tower of states that becomes light in the limit $\Lambda_d \rightarrow 0$ as
\begin{equation}
\dfrac{m_{\mathrm{tower}}}{M_P}\sim\left| \dfrac{\Lambda_d}{M_P^d} \right| ^{\gamma_d},
\end{equation}
with $\gamma_d$ a positive $\mathcal{O}(1)$ constant.
\end{tcolorbox}

	It should be clear by now that the breakdown of  gravitational EFTs by the appearance of  light towers of states is an ubiquitous feature in string theory/quantum gravity. However, this generality makes it hard to pinpoint the towers that could be more relevant to describe in our Universe. In this regard, there is particular limit which might be specially interesting, namely the one associated with the gravitino mass going to zero.\footnote{See also \cite{Palti:2020tsy} for an earlier discussion about the existence of towers associated to fermionic fields in the context of the swampland, and \cite{Antoniadis:1988jn} for a study of the existence of towers of gravitinos becoming light in the vanishing gravitino mass limit in a class of heterotic compactifications.} This is the case because the gravitino being the supersymmetric partner of the graviton, it is always there as soon as we consider any supersymmetric theory of gravity, and in fact it is generally related to the scale of spontaneous supersymmetry breaking in non-supersymmetric vacua (as the one we happen to live in). This motivated the following proposal
	\vspace{0.2cm}
	\begin{tcolorbox}[colback=boxblue]
  	\textbf{Gravitino Distance Conjecture} (GDC) \cite{Cribiori:2021gbf, Castellano:2021yye}: In a supersymmetric  theory with a non-vanishing gravitino mass $m_{3/2}$, a tower of states becomes light in the limit $m_{3/2}\rightarrow 0$ according to
		\begin{equation}
		\label{eq:GDC}
		\dfrac{m_{\mathrm{tower}}}{\mp}\, \sim \, \left( \dfrac{m_{3/2}}{\mp}\right)^\delta\,   \qquad \text{with} \quad  0<\delta \leq 1  \, .
	\end{equation}
	 
\end{tcolorbox}	
	
	One can think of the Gravitino Distance Conjecture, as a unification of the Swampland Distance Conjecture and the Anti-dS Distance Conjecture along the particular trajectories on field space selected by the vanishing of the gravitino mass. As a final comment regarding both the Anti-dS Distance Conjecture and the Gravitino Distance Conjecture, let us mention that they do not forbid the strict $\Lambda=0$ or $m_{3/2}=0$, respectively. In fact, these two situations are well-known to be realised in the string theory landscape. What these conjectures state is that this situation cannot be continuously connected with their the non-vanishing counterparts within the same EFT description, as an infinite tower of light states would kick in and spoil it. We will see in the next subsection how under certain assumptions some of these conjectures can be used to try to connect these ideas with the observable universe.

%Gravitino Distance Conjecture

\section{Implications for Particle Physics}
\label{sec:implicationsPP}

Having introduced the most relevant swampland conjectures for our purposes in section \ref{sec:conjectures}, we now present the main consequences and predictions for particle physics that have been found so far.

\subsection{Compactifying the SM: Neutrino masses, the cosmological constant and supersymmetry}
\label{ss:SMcompactifications}

Several very interesting implications for particle physics from the Swampland Program arise when the conjectures are applied to compactifications on simple manifolds, such as the circle.
If the theory compactified on these simple manifolds turns out to be inconsistent, it means that the theory itself is pathological.  

In this context, a potentially interesting situation arises if the theory upon compactification gives rise to a lower-dimensional non-supersymmetric AdS vacuum, as this would be in direct contradiction with the Non-susy AdS Conjecture, unless the vacuum is actually unstable.
 Moreover, if by compactification one can obtain a family of AdS vacua with different cosmological constant, more constraints would be obtained from imposing the AdS Distance Conjecture in the limit of vanishing cosmological constant.

It turns out to be particularly interesting to consider compactification of the Standard Model of particle physics. In fact, considering its compactification on the simplest possible space, namely a circle  (as originally done in \cite{ArkaniHamed:2007gg} for different purposes) already gives rise to very interesting restrictions when combined with the Non-susy AdS Conjecture  or the AdS Distance Conjecture, as considered in \cite{Ibanez:2017kvh,Ibanez:2017oqr,Hamada:2017yji,Gonzalo:2018tpb,Gonzalo:2018dxi} and \cite{Rudelius:2021oaz, Gonzalo:2021fma, Gonzalo:Toappear}, respectively. The way to obtain these constraints is to consider the lower-dimensional effective potential that is generated due to the Casimir effect. In the same way as vacuum fluctuations create a potential between two-parallel plates that depends on the distance between them, there is also a potential along a circular compact dimension, which depends on its radius $R$. This potential can be calculated and it takes the following asymptotic form for a massless field
\begin{equation}
\label{eq:casimirpotential}
V_{p}(R)=\pm \dfrac{n_{p}}{720 \pi} \frac{r^3}{R^{6}},
\end{equation}
where $r$ is just a constant with dimensions of length that can be fixed to any value\footnote{This $r$ is just introduced to keep the lower-dimensional metric adimensional so that the relevant component of the metric is $g_{33}=(R/r)^2$. It gives the periodicity of the coordinate in the circle, namely $y \sim y + 2 \pi r$, and the physical radius of the circular dimension is then controlled by the dimensionfull $R$, that is $2 \pi R \, = \,\int_{0}^{2\pi r} \, dy\,  \sqrt{g_{33}}\, $ .} and $n_p$ is the number of degrees of freedom of the particle. The negative sign corresponds to bosons and the positive one to fermions (with periodic boundary conditions along the circle). \footnote{For fermions with anti-periodic boundary conditions one obtains a negative contribution with a different numerical prefactor, but we will not consider it here as it is not useful to obtain phenomenological constraints, we refer the interested reader to \cite{Hamada:2017yji,Gonzalo:2018tpb} for details.} In the case of massive particles, we obtain the same leading behaviour in the limit $m\ll R^{-1}$ and an exponential suppression when $m\gg R^{-1}$, so that effectively we can ignore particles with masses above the energy scale given by $R^{-1}$. In the 3d effective potential for the ``radion" (the field associated to the radius), only massless particles will enter when $R$ is very large. As we go to smaller and smaller values for $R$, we reach the  thresholds for different particles which start to contribute to the potential until they behave effectively as massless when $R^{-1}$ is above their mass scales.

When applying this to the Standard Model compactified on a circle, we include the Einstein-Hilbert term of the action to include the gravitational sector at the EFT level, and a tiny positive cosmological constant, $\Lambda_4$. The scalar potential for the radion has then one extra contribution apart from the Casimir energies of all the particles in the spectrum, given by the 4d cosmological constant which upon compactification yields 
\begin{equation}
\label{eq:Lambdapotential}
V_{\Lambda} (R)=2 \pi r \left( \dfrac{r }{R}\right)^2 \Lambda_4,
\end{equation}
so that the full lower-dimensional potential takes the form
\begin{equation}
V_{\mathrm{tot}}(R)=V_{\Lambda}(R)+\sum_p V_p(R),
\end{equation}
where the sum over $p$ runs over the different particles in the spectrum which satisfy  \mbox{$m\ll R^{-1}$}. For very large $R$, the cosmological constant part dominates and the 3d potential is thus positive. As we go to lower values of $R$, different particles start to contribute once their mass threshold is reached and decrease or increase the potential depending on their bosonic or fermionic character, respectively. As we present momentarily, the formation of a lower-dimensional AdS vacuum (or a family of them) is highly dependent on some aspects of the spectrum of the SM, and this allows for the extraction of very interesting constraints by requiring that both the Non-susy AdS and the AdS Distance Conjecture be satisfied.

\subsubsection*{Constraints from the Non-susy AdS Conjecture}

\begin{figure}[t]
     \begin{center}
     	\subfigure[]{	
        \includegraphics[scale=0.36]{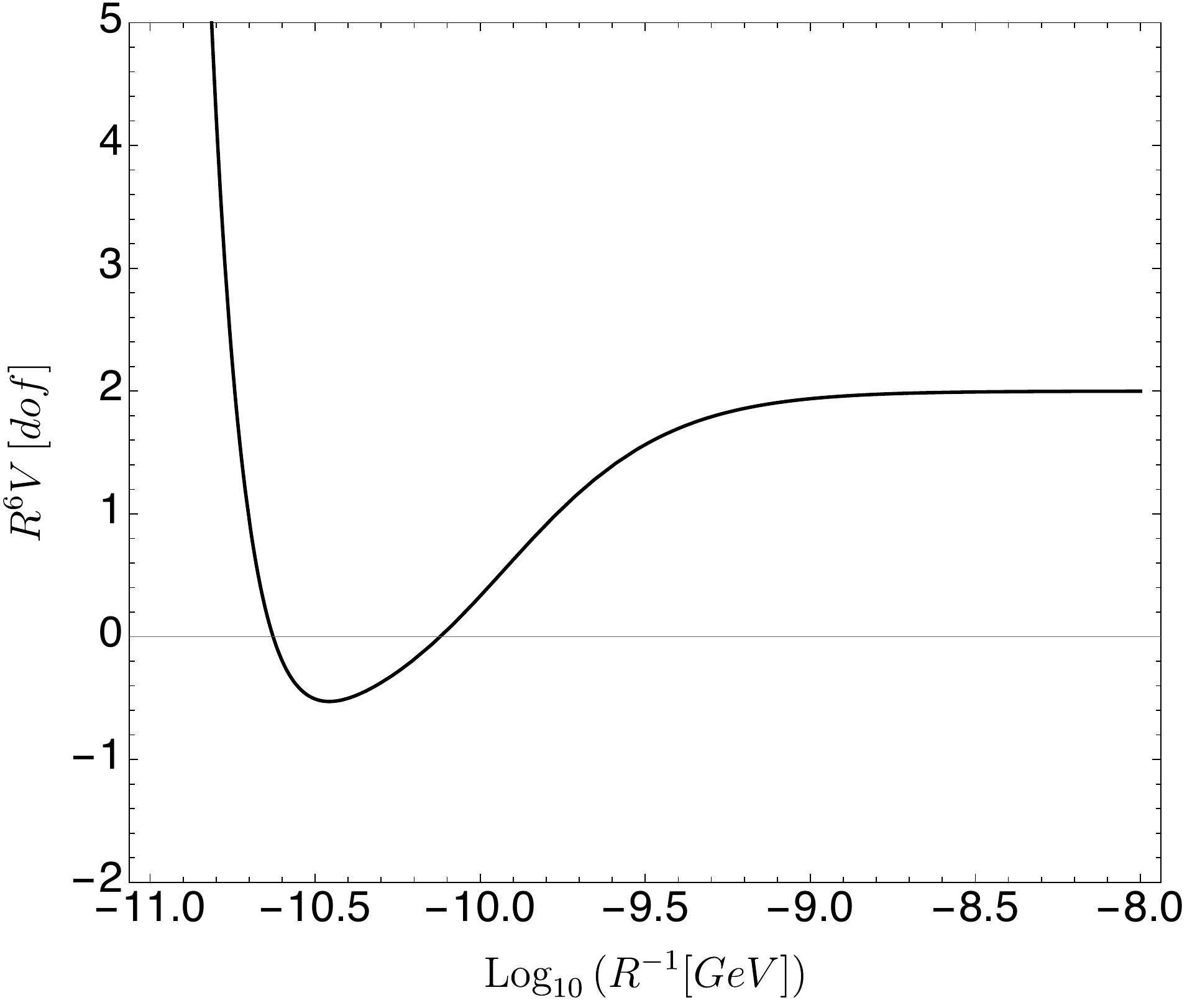}
        \label{fig:Veffmajorana}
        }
        \subfigure[]{
        \includegraphics[scale=0.355]{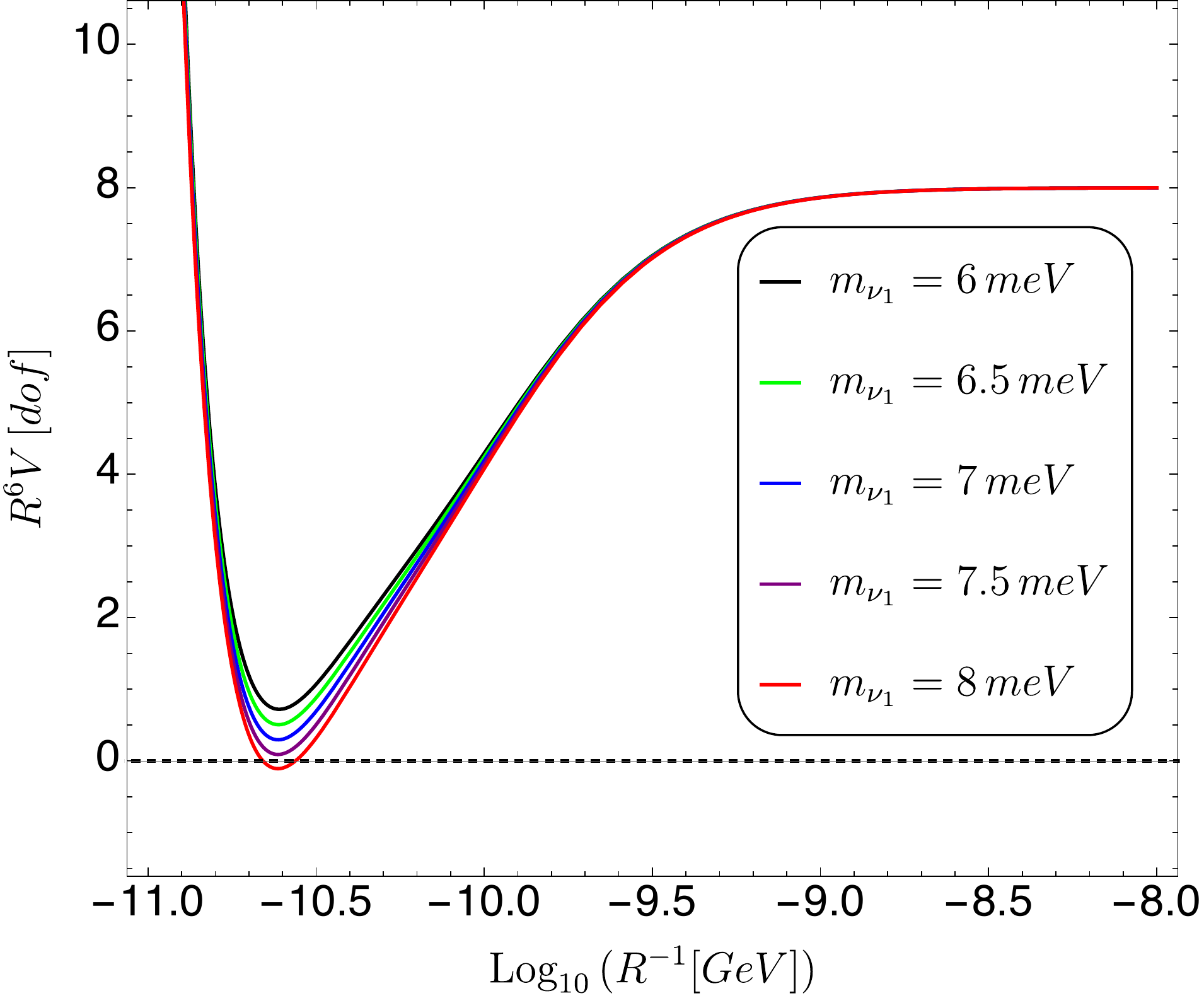}
        \label{fig:Veffdirac}
        }
      \caption{\footnotesize  Figs. taken from \cite{Gonzalo:2018tpb}. The vertical axis represents the 3d effective potential for the radion, $V(R)$ (in units of with $r=1$GeV), multiplied times $R^6$ so as to give a constant profile when $R\rightarrow 0$, and it is normalized by the contribution of one degree of freedom \textbf{(a)} Radion effective potential for Majorana neutrinos (with heavy Majorana masses) where an AdS minimum always develops. \textbf{(b)} Radion effective potential for \mbox{(pseudo-)Dirac} neutrinos with different values for the lightest neutrino mass $m_{\nu_1}$ (in mili-electron volts). The formation of an AdS vacuum can be avoided if the neutrino masses are light enough. }
      \label{fig:neutrinovacua}
      \end{center}
      \end{figure}

In going from large to small $R$, the first particles to enter the game are the massless degrees of freedom, which are the 2 degrees of freedom of the graviton and the 2 of the photon, and they decrease the potential. The next degrees of freedom that kick in are the massive (but very light) neutrinos. If they are Dirac, they contribute with 12 light degrees of freedom at low energies, whereas if they are Majorana, they contribute with something between 6 and 12 light degrees of freedom depending on the value of the Majorana masses for the right-handed neutrinos. In simple See-Saw scenarios without fine-tuning this Majorana mass is much higher than the Dirac mass so we will effectively take 6 light degrees of freedom when we talk about Majorana neutrinos, but keep in mind that whenever we refer to Dirac we should really say \mbox{pseudo-Dirac} (i.e. having both Dirac and Majorana masses of the same order) with 12 light degrees of freedom. The neutrinos would then increase the potential, and therefore an AdS minimum can be formed, as shown in Fig. \ref{fig:neutrinovacua}. If there are enough fermionic degrees of freedom in the neutrinos and if they are sufficiently light (and thus start to contribute to the potential at large enough radius) they can lift the potential before it crosses zero, otherwise an AdS vacuum will form. If this lower-dimensional AdS vacuum was stable, the Non-susy AdS Conjecture would be violated and the SM would be in the swampland. Since the Standard Model is a good low energy effective theory, unless one assumes there are additional light fermionic degrees of freedom beyond those of the Standard Model, we conclude that neutrinos must be \mbox{(pseudo-)Dirac}. Furthermore, one also obtains an upper bound for their masses in terms of the cosmological constant
\begin{equation}
\label{eq:neutrinobound}
m_{\nu}\lesssim \Lambda_{4}^{1/4}  %\sim   10^{-3} eV) \, . 
\end{equation}

This is extremely interesting as it provides an explanation for the well-known numerical coincidence between the upper bounds for neutrino masses and the value of the cosmological constant, namely $m_{\nu}^4 \sim \Lambda_{4} $ (see e.g. \cite{Planck2018, PDG2020}). Moreover, pure Majorana neutrinos with heavy Majorana masses (i.e. of See-Saw type) would be ruled out because the current experimental values for the mass difference between the neutrinos in the three generations are such that even if the lightest was massless it could never avoid the formation of the lower-dimensional AdS vacuum. This applies both for normal or inverse hierarchy.

In fact, this reasoning can also be used to relate the electro-weak scale, $v \simeq 246$ GeV,  to the cosmological constant, giving a new perspective into the electro-weak hierarchy problem. Assuming fixed Yukawa couplings the neutrino mass can be written in terms of the Higgs vev ($\left| \left\langle \Phi \right\rangle \right|=v/\sqrt{2}$) as $m_{\nu}\, =\, v \, y_{\nu} /\sqrt{2} $, so that  eq. \eqref{eq:neutrinobound} yields
\begin{equation}
\label{eq:EWccbound}
v \lesssim \dfrac{\Lambda_{4}^{1/4} }{y_{\nu}}.
\end{equation}

This being said, it is important to remark that in order for this conjecture to have some predictive power and not to lose the previous bounds, the 3d AdS vacua must be stable. For that the potential in the limit $R \rightarrow 0$ should go $V(R) \rightarrow  + \infty$, as otherwise it would not be bounded from below and AdS vacua would be unstable.  The sign of the potential in this limit is related to the $k$-th supertrace, which is defined as
\begin{equation}
\label{eq:supertrace}
\operatorname{Str}\left(M^{2 k}\right)=\sum_{b} n_{b} \, m_{b}^{2 k}-\sum_{f}  n_{f} \,  m_{f}^{2 k} \, , \quad \mathrm{with} \quad  k=0, \, 1, \, 2 \, \ldots 
\end{equation}
Here the sum over $n_b$ includes the bosonic degrees of freedom whereas the one over $n_f$ the fermionic ones. The sign of the potential in the $R\rightarrow 0$ limit goes like  \mbox{$ V_{\mathrm{tot}} \propto (-1)^{k+1} \operatorname{Str}\left(M^{2 k}\right)$} for the first non-vanishing supertrace, that is, the one with the lowest $k$ which is non-zero.  This quantity depends on the whole spectrum of the theory, and for $k=0$ is just the difference between the number of fermionic minus the number of bosonic degrees of freedom, so that for a theory in which they are not equal, the predictions are maintained if there are more fermions than bosons (as in the case of the SM). If the number of fermionic and bosonic degrees of freedom was equal, as happens in supersymmetric theories, the sign would then be determined by inserting $k=1$, and so on. Note that in an exact supersymmetric theory with all the fermions and bosons having the same masses, all the contributions would be zero, as expected. 

Going back to the SM, as one goes up in energies all the particles have to be included. Up to a small region around the QCD scale ($\sim 1$ GeV), where a perturbative description is not available, one can include all particles on the SM and due to the fact that the SM includes many more fermionic than bosonic degrees of freedom, it is guaranteed that $V(R) \rightarrow  + \infty$, at least up to scales of the order of  a few TeV.  

Considering this also allows to argue for the existence of the Higgs boson in the SM from a swampland point of view, as our SM without Higgs would be in the swampland \cite{Gonzalo:2018dxi}. Let us explain the counting of degrees of freedom that is behind this claim. If the SM had no Higgs mechanism, it would present an approximate accidental global symmetry $U(6)_L \times U(6)_R$ that is spontaneously broken to $U(6)_{L+R}$ by the QCD condensate at a scale $\Lambda_{QCD}$, yielding 36 (pseudo-)goldstone bosons. Out of these, three would be eaten to give rise to the massive $W^\pm$ and $Z$ bosons, and one gets a mass through the QCD anomaly. This would give rise to 32 light\footnote{They are expected not to be exactly massless due to electro-weak corrections, but these are numerically taken into account in \cite{Gonzalo:2018dxi} and do not qualitatively change  the picture. Light here means below $\Lambda_{QCD}$.}  bosonic degrees of freedom, which together with the 4 of the photon plus graviton add up to a total of 36 bosonic degrees of freedom. On the other hand, the number of light fermionic degrees of freedom, which come from leptons (considering \mbox{(pseudo-)Dirac} neutrinos according to the discussion above) is 24. This would cause the potential to become negative before the QCD scale and after that it is eventually dominated by fermions in the SM, so that an AdS vacuum would develop.\footnote{Incidentally, this would not be the case if the number of generations were two or one, as the approximate global symmetry for $n$ generations is $U(2n)_L \times U(2n)_R$  and it gets broken down to $U(2n)_{L+R}$. This yields $4n^2$ light bosonic degrees of freedom and $8n$ fermionic ones, which only lead to AdS vacua for $n>2$ (see \cite{Gonzalo:2018dxi} for details).} By assuming fixed Yukawa couplings one can then include a Higgs field and begin to increase its vev, $v$, slowly. Every time a quark reaches a (Higgs induced) mass of the order of $\Lambda_{QCD}$ the rank of the $U(6)$'s in the argument above is reduced by one unit, effectively lowering the number of light bosonic light degrees of freedom. Already when the heaviest quark, namely the top, reaches this point, the light bosonic degrees of freedom reduce to 25, which is almost equal to the 24 fermionic ones. Taking into account the details (see \cite{Gonzalo:2018dxi})  it turns out to be enough to avoid the AdS vacuum when the Higgs vev reaches
\begin{equation}
v\, \gtrsim \, \Lambda_{QCD} \, \sim \, 100 \  \mathrm{MeV}\, .
\end{equation}

Note that this lower bound for the electro-weak scale is independent of the bound on neutrino masses, so that it can be  combined with eq. \eqref{eq:EWccbound} to give an allowed range for the electro-weak scale in terms of the measured cosmological constant and the QCD scale, which can be applicable to our universe.

\begin{figure}[t]
	\begin{center}
		\subfigure[]{	
			\includegraphics[scale=0.355]{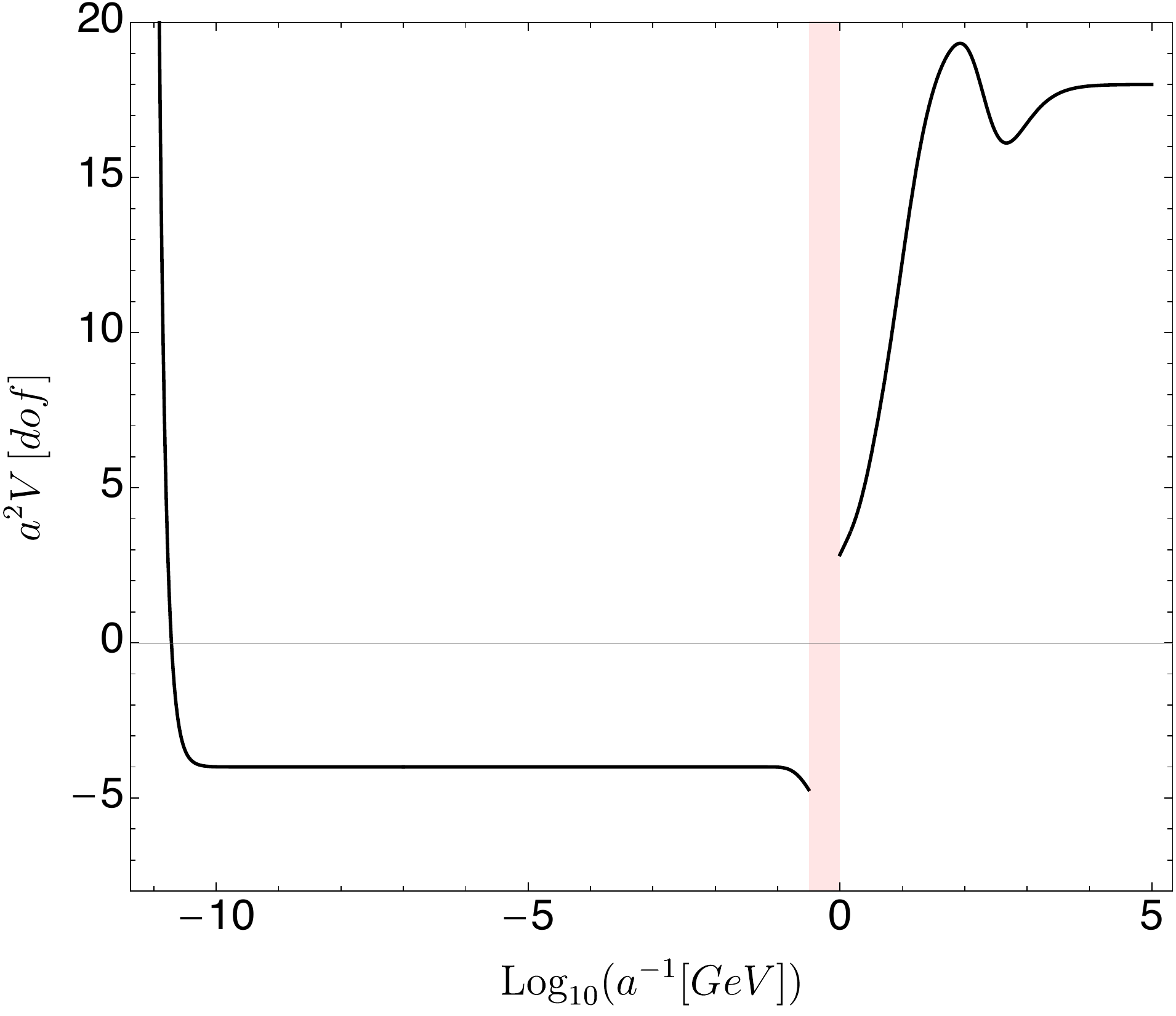} 
			\label{fig:su3SM}
		}
		\subfigure[]{
			\includegraphics[scale=0.355]{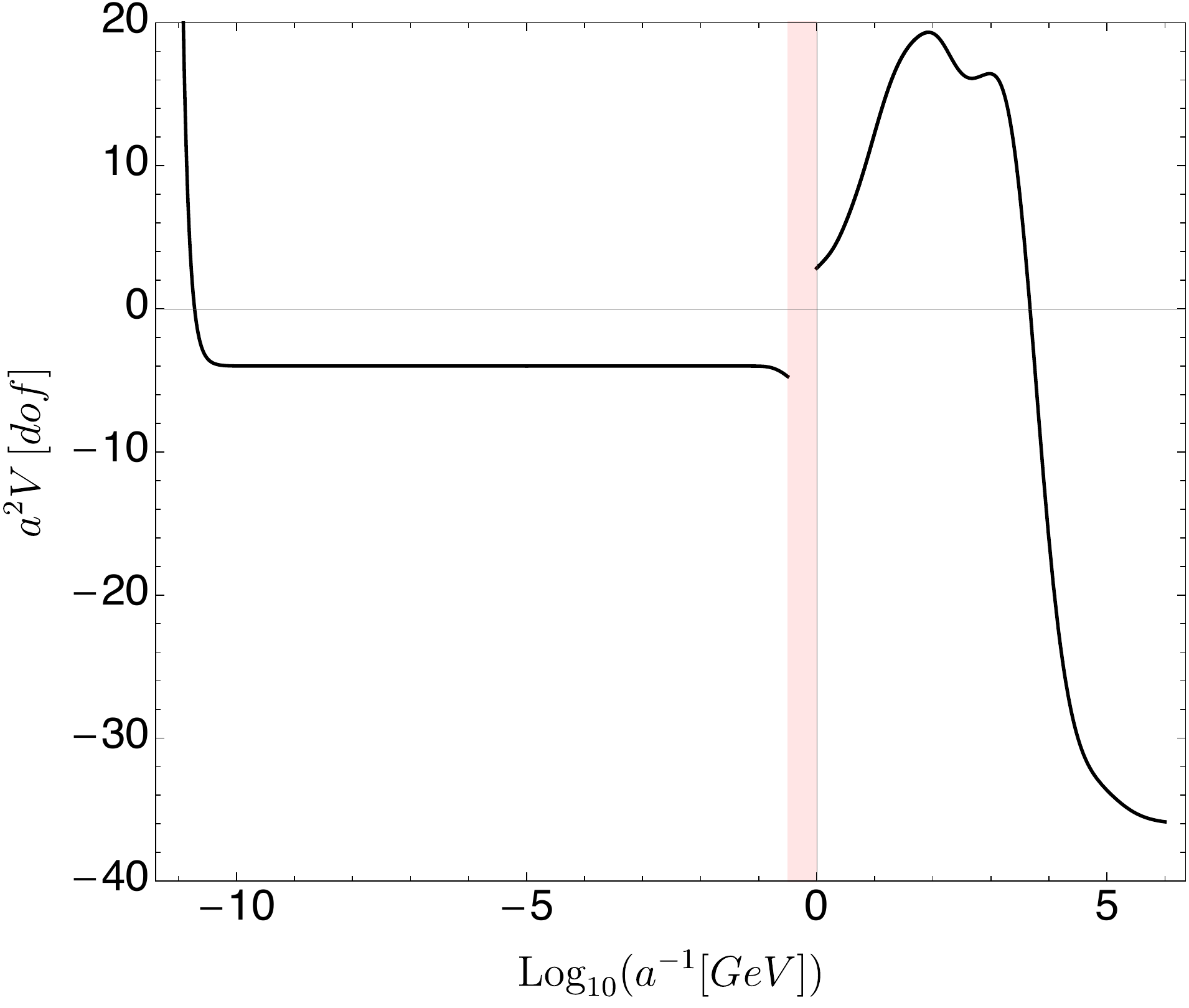}
			\label{fig:su3MSSM}
		}
		\caption{\protect \footnotesize Figs. taken from \cite{Gonzalo:2018tpb}. The red band indicates the region with energies around the QCD scale where no perturbative description is available. In analogy to the previous figures, the vertical axis shows the 3d scalar potential for the area modulus of the torus, $V(a)$, multiplied by $a^2$ to give a constant profile as $a\rightarrow 0$, and normalized by the contribution of one degree of freedom.
			\textbf{(a)}	Effective potential of the SM compactified in $T^{2}/Z_{4}$ embedded into a discrete  subgroup of the $SU(3)$ colour Cartan subalgebra, where an AdS vacuum forms just below the QCD threshold. \textbf{(b)} Same plot for the MSSM, where the extra bosonic degrees of freedom create a runaway behaviour destabilizing the AdS vacua and rescuing the SM from the swampland. }
		\label{segment_periodic_SM}
	\end{center}
\end{figure}  

At this point, one can wonder whether the SM spectrum can give rise to any other lower-dimensional AdS vacua that may constrain its parameters or the possible completions of the theory. In particular, by considering its compactification on a class of toroidal orbifolds with a particular gauge embedding (see \cite{Gonzalo:2018tpb} for details), one can study the scalar potential generated for the area modulus of the torus, which is given by $a^2$ ($a$ has units of length).\footnote{For completeness, let us mention that the analogous of eqs. \eqref{eq:casimirpotential} and \eqref{eq:Lambdapotential} for a massless particle in the torus is $V_p(a)=\pm \dfrac{n_p}{\left( 2 \pi a \right)^2}\dfrac{\mathcal{G}}{3}$, with $\mathcal{G}\simeq 0.915966$ the Catalan's constant, and $V_{\Lambda}(a)=\left( 2 \pi a \right)^2 \Lambda_4$.} In this case, the SM happens to develop another AdS vacuum which can actually not be avoided by playing with any of its free parameters, as shown in Fig. \ref{fig:su3SM}. This would actually imply that the SM alone, without any extra degrees of freedom, would be in the swampland! As explained before, the EFT that describes our universe cannot be in the swampland, and therefore this argument indicates that the SM must be extended in some way so as to avoid the formation of such an stable lower-dimensional AdS vacuum. It is important to remark at this point that this was already a well-known fact, as for example dark matter would require extra degrees of freedom, but it is still interesting to obtain it as a prediction from quantum gravity. A very economical way to fix this problem and destabilize this new AdS vacuum that forms when compactifying the SM would be to include more bosonic degrees of freedom, as they could make the potential go to $V(a) \rightarrow  - \infty$ (recall bosons give a negative contribution) as $a \rightarrow 0$. A very natural way to include bosonic degrees of freedom within a theory that has more fermionic than bosonic degrees of freedom (as the SM) is precisely supersymmetry, and it turns out that this is enough to save the SM from these new dangerous AdS vacua. In particular, a minimal susy completion of the SM would fix the problem, as shown in fig \ref{fig:su3MSSM}. At this point no preference towards any particular model (nor any susy breaking scale) appears, as long as it is supersymmetric. More contrived compactifications can be used and some restrictions may be obtained on the masses of the supersymmetric spectra, but they are very model dependent and no other generic prediction about it has been found so far.

Let us clarify an important point about the logic applied so far. The key here is that not a single vacuum of the theory must give rise to an inconsistency, or otherwise the whole theory would be in the swampland. This implies that for every dangerous compactification one has to make sure that no stable AdS vacuum appears. If the requisites needed to prevent the formation of one non-supersymmetric AdS vacuum automatically destabilized all the other would-be AdS vacua from which constraints are obtained, then these constraints would be lost, but this is not the case in the setup presented here. In particular, the constraints on neutrino masses and the cosmological constant do not disappear when the AdS vacuum that is independent of them (i.e. the one in Fig. \ref{fig:su3SM}) is cured by completing the SM with a supersymmetric spectrum. This is the case because this supersymmetric spectrum does not necessarily destabilize the neutrino AdS minimum (in particular it does not if the first non-vanishing supertrace is positive), and therefore one must still ensure that it does not form, recovering the constraints from eq. \eqref{eq:neutrinobound}.

At this point, let us also outline some of the main assumptions made to obtain these results. First of all, it is assumed that the 4d vacuum we live in is a (meta)stable dS vacuum, not a runaway  (note that this is allowed by the Trans-Planckian Censorship Censorship Conjecture \cite{Bedroya:2019snp} but not with the (refined) dS Conjecture \cite{dS1,dS2, dS3}). If it were a runaway the lower-dimensional vacuum would also have a runaway direction along the original scalar direction and therefore no lower-dimensional AdS vacua would develop. Moreover, there is the assumption that possible non-perturbative instabilities of the original 4d dS vacua do not propagate to the lower-dimensional AdS vacuum. This would be the case if the radius of the 4d bubbles mediating the instability were larger than the AdS radius of the lower-dimensional vacua, since those bubbles would contract instead of expand in lower dimensions. %Moreover, bubbles of nothing would not be a problem at the EFT level in the circle compactification as the boundary conditions for the fermions are periodic, and these are precisely the ones that do not admit such a bubble of nothing solution in circle compactifications \cite{Witten:1981gj}. 
Still, any other non-perturbative instability could destabilize the lower-dimensional vacua and spoil the predictions, even though it is not straightforward to find them. Yet an extra assumption is the fact that other scalars (arising both from scalars in the 4d theory such as the Higgs or from scalars in a string theory compactification) have not been fully taken into account. The presence of these scalars was treated in detail in \cite{Gonzalo:2018tpb}, in the context of toroidal orbifold compactifications, where most of the scalars are fixed and the same predictions are obtained. Last but not least, there is of course the assumption that the Non-susy AdS Conjecture is true, but its motivation and current status have already been summarized in section \ref{ss:non-susyAdS}. Let us remark that in spite of these possible loopholes, this setup still shows very clearly the potential implications for particle physics coming from swampland constraints, and how the swampland logic can shed some light on some apparent coincidences, such as the one between the scales of neutrino masses and the cosmological constant, or the electro-weak hierarchy problem. As we will see shortly, the fact that similar bounds can be obtained from different conjectures where the assumptions might be relaxed/changed also gives support to these ideas, even though the proofs cannot be made fully robust at present.

\subsubsection*{Constraints from the AdS Distance Conjecture}

Compactification of the SM on a circle can also yield apparent violations of the AdS Distance Conjecture introduced in section \ref{sec:towers}, as studied mainly in \cite{Gonzalo:2021fma,Gonzalo:Toappear}. Once again, insisting in the theory not to belong to the swampland may give rise to interesting phenomenological constraints. The main power of the constraints obtained here is that, unlike in the previous arguments, they do not rely on absolute stability of the lower-dimensional AdS vacua as the AdS Distance Conjecture deals with locally stable vacua, as opposed to the Non-susy AdS Conjecture. In this sense, the constraints will be independent on the behaviour of the lower-dimensional scalar potential for lower values of the radion field.

Consider the same circle compactification of the SM as before, with effective scalar potential at 1-loop given by the sum of eqs. \eqref{eq:casimirpotential} and \eqref{eq:Lambdapotential}. The idea is then to check whether the possible lower-dimensional AdS vacua belong to a family that can approach $\Lambda_{3} \rightarrow 0$, and in that case, whether a tower of states with masses going as $m_{\mathrm{tower}}\sim |\Lambda_3|^{\gamma_3}$ in Planck units. In particular, one can precisely scan different values of the neutrino masses and see that such a family of AdS vacua approaching Minkowski is realized, but the obvious tower, namely the KK tower associated to the circle, still appears as $R\rightarrow \infty$.  However,  the AdS Distance Conjecture would require a tower at finite values $R\sim 1/m_{\nu}$, which is when the Minkowski vacuum would be approached. At this point, two possibilities may arise to fix this problem. The first one is the same as before, namely that the neutrinos are \mbox{(pseudo-)Dirac} and  bounded from  above by 
\begin{equation}
\label{eq:neutrinoboundADC}
m_{\nu}\lesssim \Lambda_{4}^{1/4} \, ,
\end{equation}
so that no AdS vacuum forms in the first place and no contradiction with the conjecture appears \cite{Gonzalo:2021fma,Gonzalo:Toappear,GonzaloStrings}. Let us emphasize once more that this time it is not required that the would-be AdS vacuum be fully stable, as the potential violation of the AdS Distance Conjecture is independent of that. Hence, this argument allows for the relaxation of one of the most stringent assumptions from the previous section, namely the absolute stability of the potential lower-dimensional vacua, which could not be studied in full generality.

An alternative way out to reconcile the family of AdS vacua with decreasing cosmological constant and the AdS Distance Conjecture would be to have some correlation between the neutrino masses and the 4d cosmological constant, in such a way that $\Lambda_{3}\rightarrow 0$ could only be obtained at the same time as $R\rightarrow 0$, and therefore the KK tower would fulfil the AdS Distance Conjecture \cite{Gonzalo:2021fma,Gonzalo:Toappear,GonzaloStrings}. This kind of behaviour would happen if in the original 4d theory one had $m_{\nu}\sim |\Lambda_4|^{\gamma_4}$, which looks like a particularly big fine-running except if the neutrinos already belonged to the tower predicted by the AdS Distance Conjecture in 4d (recall that $\Lambda_4$ is small enough to be considered as $\Lambda_4 \rightarrow 0$ in the context of this conjecture). In particular, this seems to fit very naturally in the case $\gamma_4=1/4$. Thus, the second way out is that the neutrinos are the first states in the tower predicted by the AdS Distance Conjecture (applied to our 4d dS vacuum\footnote{In the original formulation of the AdS Distance Conjecture the arguments applied to argue for a tower in the limit $|\Lambda| \rightarrow 0$ are independent of the sign of the cosmological constant, so it is equally valid for dS and AdS vacua.}) and therefore in the 3d theory there would be two towers that can fulfil the conjecture. The lower-dimensional KK tower, which turns out to behave as $m_{\mathrm{KK}}\sim |\Lambda_3|^{1/3}$, and the neutrino tower, whose 3d parameter, $\gamma_3$, would depend on the 4d one, $\gamma_4$. One obtains $\gamma_3=1/3$ for the 4d  value  $\gamma_4=1/4$ in 4d, which is nothing but a particular case of the more general value $\gamma_d=1/d$, found originally in \cite{Rudelius:2021oaz} and from different arguments in \cite{Gonzalo:2021fma}.  In fact,  similar constraints relating neutrino masses with the 4d cosmological constant were argued to arise in \cite{Rudelius:2021oaz} if the 4d SM lived in AdS. Still, it was suggested there that the same bound could also apply to to the dS vacuum we appear to live in, as confirmed by \cite{Gonzalo:2021fma,Gonzalo:Toappear,GonzaloStrings}.

\subsection{Supersymmetry breaking and towers of states}

We have just mentioned some arguments in favor of supersymmetry coming from compactifications of the SM. Nonetheless, these arguments are independent from the scale of supersymmetry breaking. It would be interesting to have some argument that is sensible to this scale and can tell us something about whether quantum gravity predicts some scale, but nothing concrete has been put forward yet. However, some preliminary ideas have been suggested by highlighting the special role played by the gravitino mass in the Swampland Program, and particularly its massless limit \cite{Cribiori:2021gbf, Castellano:2021yye} (see also \cite{Palti:2020tsy} for an earlier discussion about the existence of towers associated to fermionic fields and possible relations to supersymmetry breaking) . The Gravitino Distance Conjecture introduced in \eqref{eq:GDC} implies that when the gravitino mass is small compared to the Planck mass (as will be the case in all the scenarios we are going to consider), a tower of states also becomes light. 

In Minkowski vacua, or (quasi-)dS vacua with small cosmological constant (as the one we seem to live in), the gravitino mass gives the scale of supersymmetry breaking, so that one cannot arbitrarily decouple it from the UV scales associated to the tower.  The value $\delta=1$ in \eqref{eq:GDC} would imply that the mass of the states in the tower and  the gravitino mass are of the same order, so that any susy completion of the SM would be accompanied by an infinite number of degrees of freedom (typically the scale of a KK tower coming from an extra dimension of typical size given by the same scale). In the work  \cite{Castellano:2021yye} also a lower bound $\delta \geq 1/3$ was given for string compactifications from 10d to 4d. This lower bound gives an upper bound for the separation between the tower of light states and the gravitino mass, and it is informative to consider a couple of typical scenarios for the latter and check the consequences for possible completions of the SM.

\begin{itemize}
\item[$\circ$]{ Low-energy supersymmetry breaking ($m_{3/2} \sim 1$ TeV). In this case, if supersymmetry happens to be found at energies close to the ones that are currently being probed by the LHC, from $1\geq \delta \geq 1/3$one would expect a tower at the scale
\begin{equation}
 10^3\ \text{GeV} \, \lesssim \,m_{\mathrm{tower}} \, \lesssim \, 10^{13} \  \text{GeV}\, . 
 \end{equation}
 Even though this is quite a wide range, it directly rules out the popular \emph{big desert scenario}, which includes no new physics above the low energy supersymmetry breaking scale and until $\sim 10^{16}$ GeV.
  }
 
 \item[$\circ$]{Intermediate-scale supersymmetry breaking ($m_{3/2} \sim 10^{10}$ GeV). This is the minimal case if one wants to prevent the Higgs potential from being unbounded from below (given the experimental value for the Higgs mass $m_{\Phi}=125$ GeV), as restoring supersymmetry at that scale would  render the potential positive and bounded. A tower is then expected at a scale
\begin{equation}
 10^{10}\ \text{GeV} \, \lesssim \,m_{\mathrm{tower}} \, \lesssim \, 10^{16} \  \text{GeV}\, . 
 \end{equation}
  }
\end{itemize}

It is also interesting to note that the lower bounds for the $\delta$ parameter in the Gravitino Distance Conjecture were given in terms of tensions of membranes separating different vacua (see \cite{Castellano:2021yye}). These membranes couple naturally to 3-forms, which if coupled to the Higgs and an axion as proposed in \cite{Herraez:2016dxn} (and latter revisited in \cite{Giudice:2019iwl,Kaloper:2019xfj,Lee:2019efp,Dvali:2019mhn}), allow for the Higgs mass to scan different values. Applying the Weak Gravity Conjecture to such membranes would give rise to bounds for the their tensions, and therefore for the gravitino mass. These bounds could be translated to the the Gravitino Distance Conjecture tower and might also be relevant for phenomenology.

\subsection{Phenomenological implications of the Festina Lente bound}
\label{ss:FLpheno}

Given that all cosmological observations are consistent with our vacuum being dS-like, one can directly apply the Festina Lente bound \eqref{eq:FLbound} to our universe. We will not enter here in the dichotomy of whether a fully controlled dS vacuum can be obtained from string theory or not, as this is still an open question. Instead, since the Festina Lente bound is also valid in sufficiently flat quintessence models \cite{Montero:2021otb}, we will formulate the discussion in this section in terms of a positive cosmological constant but keeping in mind that it still applies if there are no dS vacua in the landscape. This been said, one can then apply the Festina Lente bound to every particle in the SM, as well as any BSM extension thereof. We will now present its main implications following \cite{Montero:2019ekk,Montero:2021otb}.

The first thing that comes to mind is to consider the gauge coupling of electromagnetism, which together with the measured value for the cosmological constant yields 
\begin{equation}
\sqrt{e M_{P} H} \sim 10^{-3} \mathrm{eV}.
\end{equation}

This bound is fulfilled by all the particles in the SM, the lightest of which is the electron, with mass $m_\mathrm{e}\simeq 0.511$ MeV. Even though one could be tempted to say that this is amply satisfied (by about 8 orders of magnitude), let us remark that the difference between the Hubble scale ($H\sim 10^{-33}$ eV) and the Planck scale ($M_P\sim 10^{27}$ eV) is of 60 orders of magnitude, so this new universal lower bound for all charged particles which is set by the geometric mean of the two quantities is fulfilled by 8 orders of magnitude as compared to 60. 
Another way to phrase this is to rewrite the Festina Lente inequality as in the second step of \eqref{eq:FLbound}, namely in terms of the cosmological constant, which yields
\begin{equation}
\dfrac{\Lambda_4}{M_P^4} \lesssim \dfrac{m_e^4}{e^2\, M_P^2}\sim 10^{-89}.
\end{equation}
As compared to the well-known value (in Planck units) $\Lambda_4/M_P^4 \sim  10^{-120}$, the Festina Lente Bound makes a good job in reducing the gap from 120 orders of magnitude to  \emph{only} 30. Moreover, as done around \eqref{eq:EWccbound}, one can relate the electro-weak scale, $v$, with the masses of the SM particles and translate the Festina Lente bound into a relation between the electro-weak scale and the cosmological constant. In particular, by applying it to the $W^{\pm}$ bosons, which have a mass $m_w=gv/2$ (with $g$ the gauge coupling of the SM $SU(2)$ group) one obtains
\begin{equation}
v^{2} \gtrsim \dfrac{ M_{P} H}{g}
\end{equation}

\begin{figure}[t]
	\begin{center}
			\includegraphics[scale=0.8]{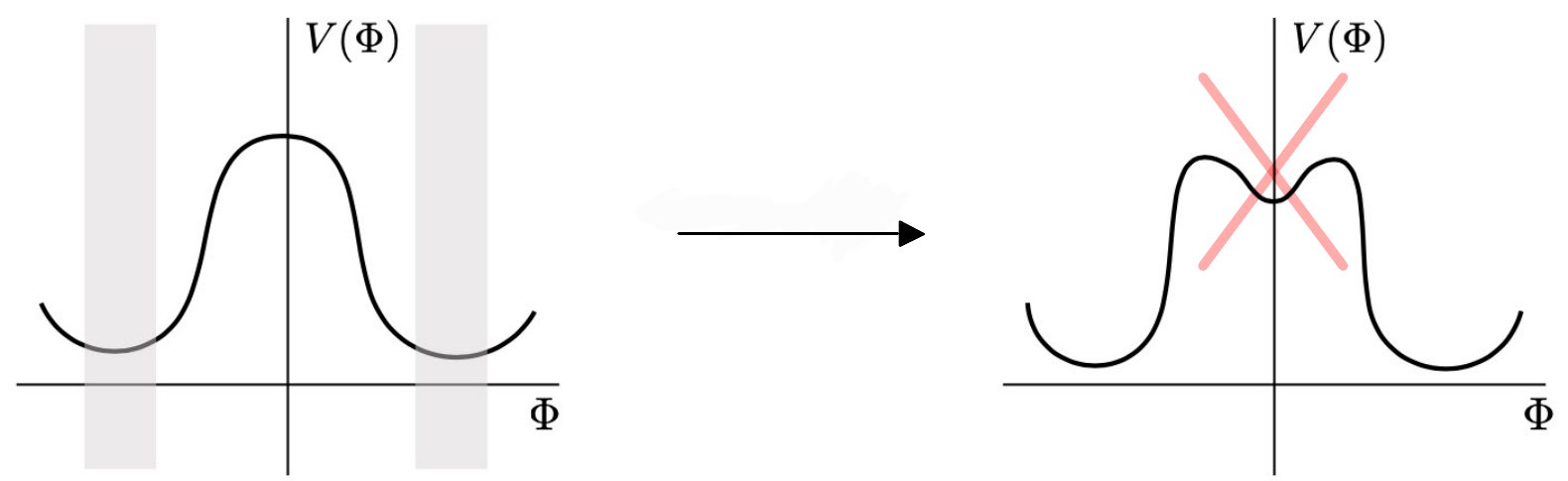} 
	\caption{\protect \footnotesize Fig. adapted from \cite{Montero:2021otb}. The \emph{Mexican hat potential} in the right, with an unstable maximum at the (symmetry restoring) point at the origin is not in tension with the Festina Lente bound. A local minimum as in the figure on the right is nevertheless in tension with the Festina Lente bound unless extreme fine-tuning is included. }
		\label{fig:Higgspotential}
	\end{center}
\end{figure}

%since the electro-weak scale, $v$, is related to the masses of the SM particles via $m_{i}\, =\, v \, y_{i} $, the Festina Lente bound can be translated to 
%\begin{equation}
%v \, y_i \, \gtrsim \, \sqrt{e M_{P} H}. 
%\end{equation}
This gives an upper bound for the vacuum energy in terms of the electro-weak scale, morally similar to \eqref{eq:EWccbound} (even though less constraining at this point). In relation to the electro-weak scale, the bound \eqref{eq:FLbound} also has some implications for the Higgs potential at the origin. In particular, it forbids the existence of any local minimum at the origin, as it would not break the electro-weak symmetry and the charged particles to which the bound can be applied would remain massless, whereas the value of the potential at such a minimum would be non-zero unless an arbitrarily high fine-tuning took place.\footnote{To be precise, this is only valid in the IR, where all the possible heavy degrees of freedom have been integrated out. More general arguments including the analysis around the UV cutoff scale are presented in \cite{Montero:2021otb}} Therefore, the Higgs potential cannot have a symmetry preserving local minimum at the origin (unless extreme fine-tuning is included) as shown in Fig. \ref{fig:Higgspotential}. Let us remark that even though the renormalizable Higgs potential has a maximum at the origin, that region has not been accessed experimentally and there is no EFT argument (nor experimental constraint) that would forbid the addition of extra non-renormalizable pieces to the potential so that such a minimum would develop.

It is also interesting to consider the situation in which non-abelian groups are present. In particular, Nariai black holes can be constructed analogously by embedding them into the corresponding Cartan subgroup. This would automatically give rise to an inconsistency due to the fact that non-Cartan gauge bosons are charged and massless. The way out is then that in any dS vacuum non-abelian gauge symmetries must be either broken or confining, with
\begin{equation}
m_{W}\gtrsim H \qquad \text{or} 	\qquad  \Lambda_{\mathrm{conf}} \gtrsim H \, , 
\end{equation}
with $m_W$ the masses of the  charged gauge bosons and $\Lambda_{\mathrm{conf}}$ the scale of confinement. This is precisely the case for the SM $SU(2)$ and $SU(3)$, respectively.

Another interesting implication from this setup comes from considering the only non-anomalous global symmetry of the SM, namely $(B-L)$. As argued in section \ref{ss:globalsymmetries}, there cannot be exact global symmetries in quantum gravity, so $(B-L)$ must be either gauged or broken at high-energies. Many BSM scenarios accommodate the possibility  that $(B-L)$ is spontaneously broken in the UV, but if instead one works with the hypothesis that it is gauged and unbroken at low energies, experimental bounds yield $e\lesssim 10^{-24}$ \cite{Cheung:2014vva,Craig:2019fdy}. Since neutrinos are charged under $(B-L)$, the Festina Lente bound would then imply that the lightest neutrino cannot be exactly massless.

Finally, let us mention that this can also lead to constraints in the parameter space of milicharged dark matter models \cite{Montero:2019ekk}. Even though these constraints are less restrictive than the current experimental constraints, meaning that these arguments cannot be used to help experimental searches at present, it is definitely a good (and essential) sign that they are compatible with the experiments.

\subsection{Massless photons}

From the EFT point of view, it is not completely ruled out that the photon can have a tiny (and technically natural) mass $m_{\gamma} \lesssim 2\times 10^{-14}$ eV (see \cite{Reece:2018zvv} and references therein), as opposed to the SM case, where it is exactly massless. This is the case because the extra longitudinal mode couples very weakly if the mass is small, making it difficult to detect, and also the renormalization of the mass does not naturally drive it to much larger values, as is the case with scalar fields. In fact, it could even seem plausible given the fact that other  parameters are known (or expected) to be small but not exactly zero (e.g. neutrino masses, the cosmological constant, the $\theta$-angle\ldots). However, as we have seen for some of the aforementioned cases, the swampland has something to say when it comes to restricting the allowed values. We present here the argument put forward in \cite{Reece:2018zvv} according to which, combining swampland reasoning with the current experimental upper bounds in the photon mass, the only possibility is an exactly massless photon.

To be more precise, this argument works for a St\"uckelberg mass for the photon, as opposed to a Higgs mass (the difference in this context will be clarified in a moment). Introducing  a St\"uckelberg field, $\theta$, the photon gets a mass through a term in the Lagrangian of the form  $\mathcal{L}_{A-\theta}=\frac{1}{2} f^{2}\left(\partial_{\mu} \theta-e A_{\mu}\right)^{2}$, where the usual gauge invariance is recovered under the transformations $A_{\mu} \rightarrow A_{\mu}+\frac{1}{e} \partial_{\mu} \alpha, \  \theta \rightarrow \theta+\alpha$. The mass of the photon is then given by $m_{\gamma}= e f$, and $f$ is the so-called axion decay constant for the St\"uckelberg field (it is called an axion because the gauge symmetry corresponds to a shift-symmetry of $\theta$). The massless limit then lies at either $e\rightarrow 0$, which is forbidden by the no-global-symmetries conjecture, or $f\rightarrow 0$. When embedded into a supersymmetric realization (as in stringy constructions), the real axion is completed  to a complex scalar field by including an extra real scalar, the saxion.  The main distinction between a Higgs mass and a St\"uckelberg mass in this context is not in the number of degrees of freedom, but in the kinetic term of the corresponding complex scalar field. It is non-standard in the St\"uckelberg case, and in fact the limit of vanishing axion decay constant, which now depends on the vev of the saxion, generically lies at infinite distance (in the sense of section \ref{sec:towers}). By applying the Swampland Distance Conjecture one then expect the breaking of the EFT due to an infinite tower of states becoming light.

To make this argument more precise, one can actually resort to the Weak Gravity Conjecture for $(p+1)$-forms introduced in eq. \eqref{eq:WGCqforms}.\footnote{This interplay between the Swampland Distance Conjecture and the (tower versions of the) Weak Gravity Conjecture is a recurring pattern in string compactifications and it has been suggested as a generic property of infinite distance points \cite{Grimm:2018ohb}, as well as used to fix order one parameters in the conjectures \cite{Gendler:2020dfp,Bastian:2020egp}.} First, one dualizes the scalar field in 4d (the axion) to a 2-form as $\frac{1}{2\pi f} \epsilon^{\mu \nu \rho}{}_{\sigma} \partial_{[\mu} B_{\nu \rho]}\, = \, f \partial_{\sigma} \theta$, which yields the kinetic term of the 2-form $\mathcal{L}_{B}=\frac{1}{12 f^{2}} H_{\mu \nu \lambda} H^{\mu \nu \lambda}$, with $H_{\mu \nu \lambda}$ the field strength of $B_{\mu \nu}$. In this description, $f$ is actually the gauge coupling of the corresponding higher-form gauge symmetry, and it is clear that the massless photon limit ($f\rightarrow 0$) corresponds to a weak coupling point. Strings are the objects that are electrically charged under the 2-form, or equivalently, magnetically charged under the dual axion, so applying then the Weak Gravity Conjecture for 2-forms to the minimally charged state yields the following bound for the tension of a string 
\begin{equation}
T \lesssim f M_P.
\end{equation}
In the $f\rightarrow 0$ limit, this gives rise to an infinite tower of states coming from the excitations of the string that becomes tensionless. This yields a cutoff scale for the EFT given by $\Lambda_{\mathrm{UV}} \simeq \sqrt{T}$, so that one obtains the bound
\begin{equation}
\Lambda_{\mathrm{UV}} \lesssim \sqrt{f M_P}\, ,
\end{equation}
which actually coincides with the cutoff scale predicted by the magnetic Weak Gravity Conjecture. The aforementioned upper bound for the photon mass can then be translated to an upper bound for $f$, given that the electromagnetic coupling $e\simeq 0.3$ is also known, and it would yield the following bound for the UV cutoff scale
\begin{equation}
\Lambda_{\mathrm{UV}} \lesssim \sqrt{\dfrac{m_{\gamma} M_P}{e}} \simeq10 \ \text{MeV}.
\end{equation}
This is clearly in contradiction with observations, implying that the photon must then be massless. As a final comment, this tower of string states can also be understood from the point of view of the Swampland Distance Conjecture. These so-called axionic strings have been argued to give rise to a flow of the saxion that diverges as the center of the string is approached, giving rise to an infinite distance in field space, with the corresponding tower of string states becoming light \cite{Lanza:2020qmt, Lanza:2021qsu}.\footnote{As opposed to this, in the Higgs case there are typically semiclassical strings at whose core the Higgs vev goes to zero (i.e. the value for which the symmetry is not broken). This difference is the reason why the claim in \cite{Reece:2018zvv} is only about St\"uckelberg masses.} 

There are of course several loopholes to this argument, let us mention some of them. For example, if the electron was not the minimally charged particle under the electromagnetic $U(1)$, the cutoff scale could be raised as the value of $e$ would be lowered. Still, this seems unlikely. Similarly, if the string that fulfils the Weak Gravity Conjecture is not minimally charged, something similar can happen, even though a big change in the expression from the UV cutoff is not expected from consistency with the magnetic Weak Gravity Conjecture. Moreover, if one considers (sub)lattice or tower versions of the Weak Gravity Conjecture this must be fulfilled for the minimally charged string.
In spite of these loopholes, this argument clearly shows how the swampland logic can lead to interesting phenomenological predictions, and identification of the possible loopholes can in fact be a useful way to point towards interesting directions in model building. Along this direction, let us finally mention that this same argument can be applied to reduce the parameter space of the so-called dark photon models (see \cite{Reece:2018zvv} for details).

\subsection{Constraints on the gauge groups}
Anomaly cancelation together with cobordism conjecture have been used in  \cite{Montero:2020icj} to constrain the rank of the gauge groups that can appear, as well as the groups themselves. It was shown for example that the rank of the gauge groups in 9d theories with sixteen supercharges has to be a multiple of eight plus one, namely
\be
r=1 \, {\rm mod} \, 8 \quad {\rm in} \ d=9 \ .
\ee
This multiplicity of 8 comes purely from cobordism conjecture: 9d theories have fermions in what is called Pin$^-$ structure, and the cobordism class of 2-manifolds is ${\mathbb Z}_8$. Triviality of the cobordism class of quantum gravity theories force this to be the boundary of a 3-manifold, and there should be eight copies of these  
cobordism defects called ``I-fold" (inversion-fold). These defects have associated moduli, which come always in multiples of 8. An anomaly-cancelation condition on the 6d gauge theory after compactification on the 3-manifold then tell us that the rank inherits this periodicity. 

Similar considerations in 8d using the cobordism class of 5-manifolds, ${\mathbf Z}_{16}$ yields
\be
r=2 \, {\rm mod} \, 8 \quad {\rm in} \ d=8 \ .
\ee
Gauge groups with these ranks are all realized in string theory, giving support to string universality, or string lamppost principle, which postulates that any gauge group in a theory of quantum gravity is realized in string theory. In 7d, however, the power of triviality of cobordism  is much less constraining in 7d, where it rules out only even ranks. We expect though that more swampland constraints will be uncovered in the near future.  

One can also constrain the dimension of non-abelian gauge groups that admit only real representations\footnote{This comes about because the I-fold is the fixed plane of a symmetry that involves charge conjugation.}, such that
\be
{\rm dim}(G) + {\rm rank}(G)=0 \, {\rm mod} \, 8 .
\ee

Again using a mixture of cobordism conjecture plus anomaly cancelation, it has recently been shown that one cannot get $G_2$ gauge groups in 8d half-maximal supersymmetric theories \cite{Hamada:2021bbz}.

\section{Summary and final comments}
\label{sec:discussion}

%The key lesson from the swampland is that not everything is compatible with quantum gravity. Apparently consistent low-energy EFTs can give rise to problems when gravity is included. This can give rise to apparently unexpected constraints on EFTs and as we have reviewed here these can be relevant for phenomenology. 

The goal of this review has been to present the main implications and results of the Swampland Program for particle physics so far.  We have begun by introducing the subset of swampland conjectures on which the constraints for particle physics are based, instead of giving a complete overview (we refer the interested reader to the nice reviews\cite{Brennan:2017rbf,Palti:2019pca, vanBeest:2021lhn}). Still, to give a logical treatment, we started from the absence of exact global symmetries \cite{Banks:2010zn} (and its generalization to trivial cobordism groups \cite{McNamara:2019rup}) in quantum gravity, and continued by introducing the Weak Gravity Conjecture \cite{ArkaniHamed:2006dz}, which can morally be seen as a refinement of the latter. From there, the absence of non-supersymmetric, stable, AdS vacua in the landscape was also argued \cite{Ooguri:2016pdq}, together with the Festina Lente bound, which was originally motivated by black hole decay arguments similar to the ones behind the Weak Gravity Conjecture but in dS space \cite{Montero:2019ekk, Montero:2021otb}. We ended the review of relevant swampland conjectures in this context by highlighting the importance of infinite towers of light states in the context of quantum gravity, originally introduced in the Swampland Program by the Swampland Distance Conjecture \cite{Ooguri:2006in} and extended later with the AdS Distance Conjecture \cite{Lust:2019zwm} and the more recent Gravitino Distance Conjecture \cite{Cribiori:2021gbf, Castellano:2021yye}.

The Swampland Program has also been the source of several new ideas in cosmology recently (see e.g. \cite{Agrawal:2018own, Cicoli:2018kdo, Agrawal:2019dlm,Bedroya:2019tba,Agrawal:2020xek}), and in general it's applications to that field are ubiquitous. We have not covered them here, but the conjectures that have triggered most of the swampland discussions in cosmology in the past few years are the (refined) dS Conjecture \cite{dS1, dS2,dS3} and the Transplanckian Censorship Conjecture (TCC) \cite{Bedroya:2019snp}, that we mention for completeness. Still, it is a fact that a sharp boundary between particle physics and cosmology cannot be drawn, so we have of course tangentially touched upon some related topic, such as the cosmological constant problem, but focusing mainly on the particle physics side. 

The main potential implications from the reviewed swampland conjectures for particle physics have been presented above and we summarize the main results here:
\begin{itemize}
\item[$\circ$]{From consistency of compactifications of the SM with the Non-susy AdS conjecture, it has been argued that pure Majorana neutrinos with large Majorana masses (as in simple See-Saw models) are inconsistent with quantum gravity, leaving \mbox{(pseudo-)Dirac} neutrinos as the only option, with an upper bound on their mass given by the cosmological constant $
m_{\nu}\lesssim \Lambda_{4}^{1/4}  \sim   10^{-3} eV \, ,  $
as argued in \cite{Ooguri:2018wrx, Ibanez:2017kvh,Hamada:2017yji}. This applies independently of whether normal or inverse hierarchy are realized.  Also some new insights into the electro-weak hierarchy problem can be obtained by  translating the upper bound for neutrino masses into an upper bound the electro-weak scale in terms of the cosmological constant \cite{Ibanez:2017oqr}, as displayed in eq. \eqref{eq:EWccbound}. 
}
\item[$\circ$]{By considering different compactifications of the SM, supersymmetry (with no preferred supersymmetry breaking scale) is favored from requiring the destabilization of lower-dimensional AdS vacua \cite{Gonzalo:2018dxi}. Also the Higgs vev can be related to the QCD scale as $v \gtrsim \Lambda_{Q C D} \sim 100 \mathrm{MeV}$ \cite{Gonzalo:2018tpb}.
}

\item[$\circ$]{The same upper bound for Dirac neutrino masses in terms of the cosmological constant is obtained by requiring consistency of compactifications of the SM with the AdS Distance Conjecture  \cite{Gonzalo:2021fma,Gonzalo:Toappear, GonzaloStrings}. A possible alternative would be that the neutrinos were the light states of a tower already in 4d with $m_{\nu}\sim |\Lambda_4|^{1/4}$ \cite{Gonzalo:2021fma,Gonzalo:Toappear,GonzaloStrings, Rudelius:2021oaz}
}

\item[$\circ$]{Preliminary results from the Gravitino Distance Conjecture suggest that low energy supersymmetry is incompatible with the \textit{big desert scenario}, as a tower with scale $m_{\mathrm{tower}}\lesssim 10^{13}$ GeV is predicted. Additionally, intermediate scale supersymmetry would require $m_{\mathrm{tower}}\lesssim 10^{16}$ GeV \cite{Castellano:2021yye}.
}

\item[$\circ$]{The Festina Lente bound applied to the SM electromagnetic $U(1)$ is satisfied by all particles in the SM \cite{Montero:2019ekk} and it gives some insight into the cosmological constant problem by reducing the well-known 120 orders of magnitude between the cosmological constant an the Planck scale  to $\Lambda_{4}\lesssim 10^{-89} M_P^4$ \cite{Montero:2021otb}. It also gives a lower bound for the electro-weak hierarchy in terms of the Hubble constant $v^2\, \gtrsim \, M_P \, H / g$ and forbids a local symmetry-preserving minimum at the origin of the Higgs potential unless extreme fine-tuning is implemented \cite{Montero:2021otb}.
}

\item[$\circ$]{Additionally, when applied to non-abelian groups, the Festina Lente reasoning gives lower bounds for the masses of massive vector bosons and  confinement scales in terms of the Hubble constant $m_{W}, \ \Lambda_{\mathrm{conf}} \, \gtrsim H$ \cite{Montero:2021otb}. 
}

\item[$\circ$]{The Weak Gravity Conjecture for strings which are magnetically charged under axions giving rise to St\"uckelberg masses for photons allows to argue in favour of the SM photon being exactly massless. Otherwise a UV cutoff scale $\Lambda_{\mathrm{UV}} \lesssim   10 \mathrm{MeV}$ would be predicted, which is incompatible with observations \cite{Reece:2018zvv}.
}

\item[$\circ$]{Combining triviality of cobordism conjecture with anomaly cancelation arguments, one gets allowed ranks of gauge groups, some forbidden groups as well as relations between the dimension and the rank \cite{Montero:2020icj,Hamada:2021bbz}.}

\item[$\circ$]{Finally, even though we have not discussed the (refined) \textit{dS Conjecture} \cite{dS1, dS2,dS3} here, there are particularly remarkable implications from applying it to the SM QCD vacuum \cite{March-Russell:2020lkq}. In particular, for fixed Yukawa couplings, the extrapolation of large $N$ results to $N=3$ suggest that $v\lesssim 50$ TeV is needed to avoid the formation of metastable dS vacua, even though full lattice computations have not been able to address the formation of these metastable states yet.  
}
\end{itemize}

As a closing remark, let us emphasize that the notion of \emph{naturalness} that arises in the context of EFTs seems to be reformulated in the context of the swampland, which gives rise to new relations between apparently disconnected scales. These relations may seem obscure at the moment, but this is arguably a consequence of the fact that many of the underlying reasons behind several swampland conjectures are still not fully understood in detail. Still, it is specially encouraging how the drastic reduction of the \textit{a priori} allowed parameter space that appears when swampland conjectures are applied happens to be consistent with experimental observations so far, and to give rise to some new predictions. Even though the phenomenological constraints from the swampland are still beginning to be explored, many new interesting insights have already been found and the expectation is that these will keep growing, so that a true bridge between quantum gravity and Phenomenology can arise from the swampland conjectures. Hopefully, progress along different lines of research (both more formal and more phenomenological) can shed some light into the underlying principles of quantum gravity and help us uncover the fundamental laws of nature.

%Maybe mention that we have not covered cosmology explicitly but there are a lot of applications there too...

%Outlook
 \vspace{5mm}
{\bf \large Acknowledgments}

\noindent We are deeply grateful to our collaborators in this area of investigation, especially Alberto Castellano, Anamar\'ia Font, Eduardo Gonzalo and Luis Ib\'a\~nez. We are also grateful to Eduardo Gonzalo, Luis Iba\~nez, Miguel Montero, Salvador Rosauro-Alcaraz, and Raquel Santos-Garc\'ia for useful discussions and comments on the draft. This work was mainly supported by the ERC Consolidator Grant 772408-Stringlandscape.

\bigskip

\noindent

%We are deeply grateful to our collaborators in this area of investigation, and especially to .... This work has been supported by ...

\fontsize{11}{12}\selectfont

\bibliography{refs-review}

\bibliographystyle{utphysmodb}

\end{document}